\documentclass[ALICE,manyauthors]{cernphprep}

\usepackage[comma,square,numbers,sort&compress]{natbib}
\usepackage{hyperref}
\usepackage{lineno}
\usepackage{epstopdf}
\usepackage{epsfig}
\usepackage{color}
\usepackage[normalem]{ulem}
\usepackage{subfloat}

\newcommand{\pPb}{$p$-Pb at $\sqrt{s_{\rm NN}} = 5.02$~TeV }
\newcommand{\Pb}{Pb-Pb at $\sqrt{s_{\rm NN}} = 2.76$~TeV }
\newcommand{\qt}{$c_{n}\{2\}$ }

\newcommand{\qn}{$c_{n}\{m\}$ }

\begin{document}%

\begin{titlepage}
\PHyear{2014}
\PHnumber{105}      
\PHdate{20 May}  
%

\title{Multi-particle azimuthal correlations in p--Pb and Pb--Pb collisions\\at the CERN Large Hadron Collider}
\ShortTitle{Multi-particle azimuthal correlations in p--Pb and Pb--Pb collisions}   

\Collaboration{ALICE Collaboration\thanks{See Appendix~\ref{app:collab} for the list of collaboration members}}
\ShortAuthor{ALICE Collaboration} 

\begin{abstract}
Measurements of multi-particle azimuthal correlations (cumulants) for charged particles in \pPb and \Pb collisions are presented. They help address the question of whether there is evidence for global, flow-like, azimuthal correlations in the $p$-Pb system. Comparisons are made to measurements from the larger Pb-Pb system, where such evidence is established. In particular, the second harmonic two-particle cumulants are found to decrease with multiplicity, characteristic of a dominance of few-particle correlations in $p$-Pb collisions. However, when a $|\Delta \eta|$ gap is placed to suppress such correlations, the two-particle cumulants begin to rise at high-multiplicity, indicating the presence of global azimuthal correlations. The Pb-Pb values are higher than the $p$-Pb values at similar multiplicities. In both systems, the second harmonic four-particle cumulants exhibit a transition from positive to negative values when the multiplicity increases. The negative values allow for a measurement of $v_{2}\{4\}$ to be made, which is found to be higher in Pb-Pb collisions at similar multiplicities. The second harmonic six-particle cumulants are also found to be higher in Pb-Pb collisions. In Pb-Pb collisions, we generally find $v_{2}\{4\} \simeq v_{2}\{6\}\neq 0$ which is indicative of a Bessel-Gaussian function for the $v_{2}$ distribution. For very high-multiplicity Pb-Pb collisions, we observe that the four- and six-particle cumulants become consistent with 0. Finally, third harmonic two-particle cumulants in $p$-Pb and Pb-Pb are measured. These are found to be similar for overlapping multiplicities, when a $|\Delta\eta| > 1.4$ gap is placed.
\end{abstract}
\end{titlepage}
\setcounter{page}{2}

\section{Introduction}

The primary goal of studies with relativistic heavy-ion collisions is to create the quark gluon plasma (QGP), a unique state of matter where quarks and gluons can move freely over large volumes in comparison to the typical size of a hadron. Studies of azimuthal anisotropy for produced particles have contributed significantly to the characterization of the system created in heavy-ion collisions. These studies are based on a Fourier expansion of the azimuthal distribution given by \cite{Voloshin:1994mz}:
\begin{equation} 
\label{eq:v_n}
\frac{dN}{d\varphi} \propto 1 + 2\sum_{n=1}^{\infty} v_{n} \cos [n(\varphi-\Psi_{n})],
\end{equation}  
where $\varphi$ is the azimuthal angle of produced particles. In heavy-ion collisions, the $v_{n}$ terms generally represent flow coefficients where $n$ is the flow harmonic and $\Psi_n$ is the corresponding flow angle. The flow coefficients are believed to reflect the response of the system to spatial anisotropies in the initial state. Measurements of the second harmonic flow coefficient ($v_{2}$, elliptic flow) received keen attention at Relativistic Heavy Ion Collider (RHIC), where the correspondence with hydrodynamic calculations in Au+Au $\sqrt{s_{\rm NN}} = 200$ GeV collisions indicated that an almost perfect liquid had been produced in the laboratory \cite{Back:2004je, Arsene:2004fa, Adcox:2004mh, Adams:2005dq}. Larger values of integrated $v_{2}$ have been observed at the Large Hadron Collider (LHC) in \Pb collisions, indicating that the system created at this new energy regime still behaves as an almost ideal liquid \cite{Aamodt:2010pa}. While the initial state anisotropy is usually dominated by an elliptical overlap area which gives rise to $v_{2}$, measurements of the third harmonic flow ($v_{3}$, triangular flow) demonstrated initial state fluctuations modulate the overlap area, and they provide additional constraints to the transport coefficients of the system (e.g. the value of the shear viscosity over entropy ratio $\eta/s$) \cite{ALICE:2011ab, Aamodt:2011by, ATLAS:2012at, Chatrchyan:2012wg, Adamczyk:2013waa}. The combination of the second and higher harmonic flow coefficients manifest themselves in two-particle correlation structures (along  $\Delta \eta$) such as the away-side double hump ($\Delta \varphi \sim \pi$), and near-side ridge ($\Delta \varphi \sim 0$) observed both at RHIC and the LHC. 

The study of $p$-Pb collisions, which usually provides baseline measurements for the quantification of cold nuclear matter effects, led to a number of unexpected results \cite{CMS:2012qk, Abelev:2012ola, Aad:2012gla, Aad:2013fja, Chatrchyan:2013nka, Abelev:2012di, Abelev:2013haa}. The CMS Collaboration reported the development of a near-side ridge-like structure in high-multiplicity $p$-Pb collisions \cite{CMS:2012qk, Chatrchyan:2013nka}. We discovered a symmetric double ridge structure on both the near- and the away-side after subtracting from the high-multiplicity $p$-Pb correlation function the dominant jet contribution using the low multiplicity events \cite{Abelev:2012ola}. The ATLAS Collaboration confirmed the appearance of such structure using a similar subtraction technique \cite{Aad:2012gla}. We extended the measurements to identified hadrons and reported a mass ordering in the $p_{\rm T}$ differential $v_2$ measurements for the different species, with a crossing of $p$ and $\pi$ $v_2$ at large $p_{\rm T}$ \cite{Abelev:2012di}. Around a similar time, the CMS and ATLAS Collaborations measured finite values of $v_2$ from four particle correlations.\cite{Aad:2013fja, Chatrchyan:2013nka}. 

The origin of the ridge structure in $p$-Pb collisions has been the subject of speculation within the heavy-ion community \cite{Bozek:2012gr, Dusling:2012wy, Dumitru:2013tja, Basar:2013hea}. It has been suggested that a high enough energy density is achieved in \pPb collisions to induce hydrodynamic flow using a Lattice QCD equation of state \cite{Bozek:2012gr}. Combined with spatial anisotropies in the initial $p$-Pb state, this mechanism would induce global correlations of soft particles with significant values of $v_{2}$ and $v_{3}$. A second proposal is that the ridge arises from collimated (in $\Delta \varphi$) correlated two-gluon production from the color glass condensate (CGC) \cite{Dusling:2012wy}. This leads to few-particle correlations, rather than a global modulation of soft particles. Finally, the third explanation invokes the CGC initial state with a finite number of sources that form the eccentricity \cite{Dumitru:2013tja}. In contrast to the previous explanation, this approach allows for non-zero values of $v_{2}$ from four, six, and eight particle correlations in high multiplicity $p$-Pb collisions.

Whether the current measurements in high-multiplicity $p$-Pb events reveal the onset of collective behavior, or can be explained in terms of few-particle correlations (i.e.\ non flow), is the main goal of this analysis. We report the multiplicity dependence of the two-, four-, and six-particle correlations (cumulants) for charged particles, that can be used as a tool to investigate multi-particle correlations of various harmonics \cite{Borghini:2001vi, Bilandzic:2010jr}. We present the results in both $p$-Pb and Pb-Pb collisions at $\sqrt{s_{\rm NN}} = 5.02$ TeV and $\sqrt{s_{\rm NN}} = 2.76$ TeV respectively. The multiplicity dependence of these measurements will help decipher how flow and non flow contribute. In Section II, we will introduce multi-particle cumulants and discuss their response to non flow and flow fluctuations. In Section III we will describe the analysis details. Section IV shows our results, and Section V presents our summary.

\section{Multi-particle cumulants} 

The measurements of $v_n$ in Eq. 1 can be done using a variety of methods, which have different sensitivities to flow fluctuations (event-wise variations in the flow coefficients) and non flow. non flow refers to correlations not related to the common symmetry plane $\Psi_n$, such as those due to resonances and jets. Multi-particle cumulants are utilized since their response to flow fluctuations and non flow is considered well understood. For a given harmonic $n$, the average strength of two-particle correlations is determined by forming the following from all pairs:
\begin{equation} 
\label{eq:a2}
\langle 2 \rangle = \langle  e^ {in(\varphi_{1}-\varphi_{2})} \rangle.
\end{equation}  
The $\varphi$ values used in the subtraction will originate from different particles to prevent auto-correlations. The single angular brackets denote averaging of particle pairs within the same event. The two-particle cumulant is obtained by averaging $\langle 2 \rangle$ over an event ensemble, and is denoted as:
\begin{equation} 
\label{eq:a2}
c_{n}\{2\} = \langle \langle 2 \rangle \rangle\,.
\end{equation}  
In the absence of non flow, \qt provides a measure of $\langle v_{n}^{2} \rangle$ without the need to measure $\Psi_n$. Respectively, the average strength of four particle correlations is determined by forming the following from all quadruplets: 
\begin{eqnarray} 
\label{eq:a41}
\langle 4 \rangle = \langle e^ {in(\varphi_{1}+\varphi_{2} - \varphi_{3}-\varphi_{4}  )} \rangle.
\end{eqnarray}  
Consequently, the four-particle cumulant is then:

\begin{equation} 
\label{eq:QC4}
c_{n}\{4\} = \langle \langle 4 \rangle \rangle -  2\langle \langle 2\rangle\rangle^2.
\end{equation} 
The subtraction removes non flow contributions present in two particle correlations. In the absence of non flow, $c_{n}\{4\}$ provides a measure of $\langle v_{n}^{4} \rangle-  2\langle v_{n}^{2} \rangle^{2}$. Respectively, the average strength of six particle correlations is determined by forming the following from all sextuplets:
\begin{eqnarray} 
\label{eq:a61}
\langle 6 \rangle = \langle e^ {in(\varphi_{1}+\varphi_{2} +\varphi_{3} - \varphi_{4}-\varphi_{5}-\varphi_{6} )} \rangle.
\end{eqnarray}  
The six-particle cumulant is then:
\begin{equation} 
\label{eq:QC4}
c_{n}\{6\} =  \langle \langle 6 \rangle \rangle  -9  \langle \langle 4 \rangle \rangle \langle \langle 2 \rangle \rangle + 12\langle \langle 2\rangle\rangle^3\,.
\end{equation} 
In this case, the subtraction removes non flow contributions present in two- and four-particle correlations. In the absence of non flow, $c_{n}\{6\}$ provides a measure of $\langle v_{n}^{6} \rangle - 9\langle v_{n}^{4} \rangle \langle v_{n}^{2} \rangle + 12\langle v_{n}^{2} \rangle^{3}$. As mentioned earlier, the quantities $\langle 2 \rangle$, $\langle 4 \rangle$, or $\langle 6 \rangle$ can be determined by averaging over all particles in a given event. The quantities can also be determined using the $Q$-cumulants of different harmonics, which offers a highly efficient method of evaluating multi-particle correlations without having to consider all combinations \cite{Bilandzic:2010jr}. The flow coefficients from two-, four-, and six-particle cumulants can finally be obtained from:
\begin{eqnarray} 
v_{n}\{2\} &=&\sqrt{c_{n}\{2\}}, \label{eq:v22}\\
v_{n}\{4\} &=& \sqrt[4]{-c_{n}\{4\}}, \label{eq:v24}\\
v_{n}\{6\} &=& \sqrt[6]{\frac{1}{4}c_{n}\{6\}}\,. \label{eq:v26}
\end{eqnarray} 
If the value of $v_n$ does not fluctuate and there is no non flow, $v_{n}\{2\}=v_{n}\{4\}=v_{n}\{6\}$. A variation in $v_{n}$ on an event by event basis leads to differences in each of the values. If the variation is presented with a characteristic standard deviation $\sigma_{v_n}$, $v_{n}\{2\}= \sqrt{{\langle v_{n} \rangle}^{2}+ \sigma_{v_n}^{2}}$. When $\sigma_{v_n} \ll v_n$, $v_{n}\{4\}=v_{n}\{6\}= \sqrt{{\langle v_{n} \rangle}^{2}- \sigma_{v_n}^{2}}$ \cite{Voloshin:2008dg, AnteT}. Therefore, the difference in $v_{n}\{2\}$ and $v_{n}\{4\}$ can be used to infer the scale of $v_{n}$ fluctuations $\sigma_{v_n}$. The presence of non flow influences the cumulants as follows. Assuming large multiplicity events are a superposition of low multiplicity events, the contribution from non flow (or few-particle correlations) is expected to be diluted as \cite{Voloshin:2008dg}:
\begin{equation} 
\label{eq:nfd}
c_{n}\{m\}  \propto \frac{1}{M^{m-1}},
\end{equation} 
%
%
where $M$ is the multiplicity of the event. Therefore measuring both $c_{n}\{2\}$, $c_{n}\{4\}$, and $c_{n}\{6\}$ as a function of multiplicity will help determine whether the underlying correlations are global or few-particle. One can also suppress non flow by requiring the particles to have a relatively large separation in $\eta$, since resonances and jets will produce particles with similar rapidity.

\section{Analysis details}

The two data sets analyzed were recorded during the $p$-Pb (in 2013) and the Pb-Pb (in 2010) runs at a center of mass energy of $\sqrt{s_{\rm NN}} = 5.02$~TeV and $\sqrt{s_{\rm NN}} = 2.76$~TeV, respectively. The Pb-Pb run had equal beam energies giving a nucleon-nucleon center of mass system with rapidity $y_{\rm NN} = 0$. However, the $p$-Pb run had different beam energies per nucleon for the $p$ and Pb beam, and resulted in a center of mass system moving in the laboratory frame with $y_{\rm NN}=0.465$. All kinematic variables are reported in the laboratory frame. Charged particles are detected using the time projection chamber (TPC), the primary tracking detector of ALICE. The TPC has an angular acceptance of $0 < \varphi < 2 \pi$, $|\eta| < 0.9$ for tracks with full radial track length ($\varphi$ is the azimuthal angle and $\eta$ is the pseudo-rapidity), and $|\eta| < 1.5$ for tracks of reduced length. Information from the inner tracking system (ITS) is used to improve the spatial resolution of TPC tracks, which helps with the rejection of secondary tracks (i.e.~not originating from the primary vertex). Primary vertex information is provided by the TPC and the silicon pixel detector (SPD). Two VZERO counters, each containing two arrays of 32 scintillator tiles and covering $2.8<\eta<5.1$ (VZERO-A ) and $-3.7<\eta<-1.7$ (VZERO-C), provide information for triggering and event class determination. A more detailed description of the ALICE detector can be found elsewhere \cite{Aamodt:2008zz}. 

For Pb-Pb collisions, events are selected using a minimum bias trigger, which requires a coincidence of signals in the two VZERO detectors. We use minimum bias and high-multiplicity triggers for $p$-Pb collisions. As with Pb-Pb, the $p$-Pb minimum bias trigger, requires a coincidence of two signals from the VZERO detectors, and accepts 99.2\% of the non-single diffractive cross section. The high-multiplicity trigger requires a large number of hits in the SPD. Pile-up events are rejected by removing events with multiple vertices, and ensuring the vertices reconstructed from the TPC and SPD agree within 0.8 cm. After the pile-up rejection procedures, the results are stable with respect to luminosity.  Only events with a reconstructed primary vertex within $\pm$10 cm from the center of the detector along the beam axis are used in the analysis to ensure a uniform acceptance in $\eta$. The resulting analyzed event sample consisted of about 110M $p$-Pb and 12M Pb-Pb minimum bias events. In $p$-Pb collisions, the high-multiplicity trigger allowed for a factor of 10 increase in high-multiplicity events in the top 0.014\% of the cross section, compared to the number of minimum bias events. The $p$-Pb high multiplicity events are used for the last two data points for $n=2$, and the last data point for $n=3$. Minimum bias events are used for all other points.

The tracks used to determine the cumulants have kinematic cuts $0.2 < p_T < 3$ GeV/$c$ and $|\eta| < 1$. The tracks use an SPD hit if one exists within the trajectory, if not, they are constrained to the primary vertex. Such a configuration leads to a flat $\varphi$ acceptance. It was found that residual non-uniformities influence the cumulant extraction at a level of less than 0.1\%. We therefore do not apply acceptance corrections. Track quality is ensured by requiring tracks to have at least 70 TPC clusters out of a maximum of 159, and a $\chi^2$ per TPC cluster less than 4 for the track fit. In addition, the distances of closest approach to the primary vertex in the $xy$ plane and $z$ direction are required to be less than 2.4 cm and 3.2 cm respectively \cite{Abelev:2012ej}.

The results in this article are reported as a function of the corrected multiplicity, $\langle N_{\rm ch} \rangle$. The multiplicity corresponds to the number of charged tracks with $0.2 < p_T < 3$ GeV/$c$ and $|\eta| < 1$, corrected for tracking efficiencies. The tracking efficiency is calculated from a procedure using HIJNG (Pb-Pb) or DPMJET ($p$-Pb) events \cite{Wang:1991hta, Roesler:2000he}. GEANT3 is used for transporting simulated particles, followed by a full calculation of the detector response (including production of secondary particles) and track reconstruction done with the ALICE simulation and reconstruction framework \cite{Brun:1994aa, ALICE:2005aa}. The tracking efficiency is $\sim 70 \%$ at $p_{T}\sim0.2$ GeV/$c$ and increases to an approximately constant value of $\sim 80 \%$ for $p_{T} > 1 $ GeV/$c$. There are differences on the order of a few percent when comparing between the two collision systems due to the change in detector performance between each run. The final number of particles ($\langle N_{\rm ch} \rangle$) is extracted by correcting the raw transverse momentum spectrum with the $p_{\rm T}$ dependent tracking efficiencies. Tables \ref{t2} and \ref{t3} (which are in the Appendix) show multiplicities for the two systems and the fractional cross section.

To reduce the influence of the tracking efficiency on the cumulants ($c_n\{m\}$), we flatten the $p_{\rm T}$ dependent efficiencies by randomly rejecting high $p_{\rm T}$ particles. These particles have slightly larger efficiencies compared to the low $p_{\rm T}$ ones, so the procedure effectively re-weights the cumulants in favour of low $p_{\rm T}$ particles. This decreases the integrated value of $v_{n}$ by roughly 3\%, since $v_{n}$ generally increases with $p_{\rm T}$. Regarding the choice of multiplicity bin size, it was previously realized that event by event multiplicity fluctuations within a class having a wide multiplicity range can bias the measurement of $c_n\{4\}$, particularly in the low multiplicity region \cite{AnteT, Chatrchyan:2013nka}. We avoid this by first extracting \qn in unit multiplicity bins (i.e. $N_{\rm ch} = 6, 7, 8...$). The number of combinations scheme \cite{Bilandzic:2010jr}, or simple unit event weights gives the same values of \qn for unit multiplicity bins. We then average those values to produce \qn for larger bin widths, which have a better statistical precision. The following relation is used for averaging procedure; $\langle  y \rangle = \frac {\sum_{i} w_{i} y_{i}}{\sum_{i} w_{i}}$, where $y_i$ is the value of the cumulant in a single multiplicity bin, $w_{i}$ corresponds to a choice of weight, and $\langle  y \rangle$ is the average value obtained from the number of bins in the sum. Monte Carlo studies with known probability density functions (p.d.f.) show that when using unit weights (i.e. $w_{i}=1$), our result lies within $<0.1\%$ from the known input $\langle y \rangle$ (from the p.d.f.). Other weighting schemes such as $w_{i}=M$, where $M$ is the multiplicity of the event, or $w_{i} =1/\sigma_{i}^{2}$ where $\sigma_{i}$ is the statistical uncertainty of the bin, gave differences of around 2\%.

Additional sources of systematic uncertainties in the calculation of \qn were extracted by varying the closest approach to the vertex for the tracks, the cut on the minimum number of TPC clusters, the position of the primary vertex and, finally, by analyzing the event sample separately according to the orientation of the magnetic field. 

We also generated events with the AMPT model \cite{Lin:2004en} (which includes flow correlations) that were used as an input to our reconstruction simulations. The cumulants obtained directly from the model were compared to those from reconstructed tracks. We found small differences, which are part of the systematic uncertainties. Table \ref{tab:sysPb} summarizes the systematic uncertainties for each collision system. The final systematic uncertainty is calculated by adding all the individual contributions in quadrature. In the Appendix, Tables \ref{t2} and \ref{t3} show the multiplicities for the two systems and the fractional cross section.

\begin{table}[h]
\begin{center}
\begin{tabular}{c|c|c|c|c}
\hline
{ \bf $p$-Pb source }  & {\bf $c_{2}\{2\}$ } & {\bf $c_{2}\{4\}$ } & { \bf $c_{2}\{6\}$ } & { \bf $c_{3}\{2\}$} \\ 
\hline
Primary vertex position & $ 0.3\%$ &   n/a & n/a &  $0.7\%$ \\ 
Track type & $ 2.2\%$ & $4.0\%$ & $6.0\%$ & $2.6\%$ \\ 
No. TPC clusters & $0.2 \%$ & n/a & n/a &$0.2 \%$ \\ 
Comparison to Monte Carlo & $1.7 \%$ &  $2.9 \%$ & $4.5 \%$ & $3.3 \%$ \\ 
{ \bf Total }  & $2.8 \%$ & $4.9\%$ & $7.5\%$ & $4.3 \%$ \\ 
\hline
\hline
{ \bf Pb-Pb source }  & {\bf $c_{2}\{2\}$ } & {\bf $c_{2}\{4\}$ } & { \bf $c_{2}\{6\}$ } & { \bf $c_{3}\{2\}$} \\ 
\hline
Primary vertex position  & 0.5\% & n/a & n/a &  n/a \\
Track type  & $ 2.9\%$ & $ 6.1\%$ & $9.1\%$ &  $4.0\%$ \\ 
Sign of B-field  & 0.2\% & n/a & n/a & 0.2\% \\
Comparison to Monte Carlo & $1.7 \%$ &  $2.9 \%$ & $4.5 \%$ &$3.3 \%$\\ 
{ \bf Total }  & $3.9 \%$ & $6.8\%$ & $10.2\%$ & $5.2 \%$ \\ 
\hline
\end{tabular}
\end{center}
\caption{Summary of systematic uncertainties for $p$-Pb and Pb-Pb collisions (the acronym n/a stands for non applicable).}
\label{tab:sysPb}
\end{table}

\newpage
\section{Results}

\subsection{The second harmonic two-particle cumulant}

\begin{figure}[th!f]
\includegraphics[width = 1\textwidth]{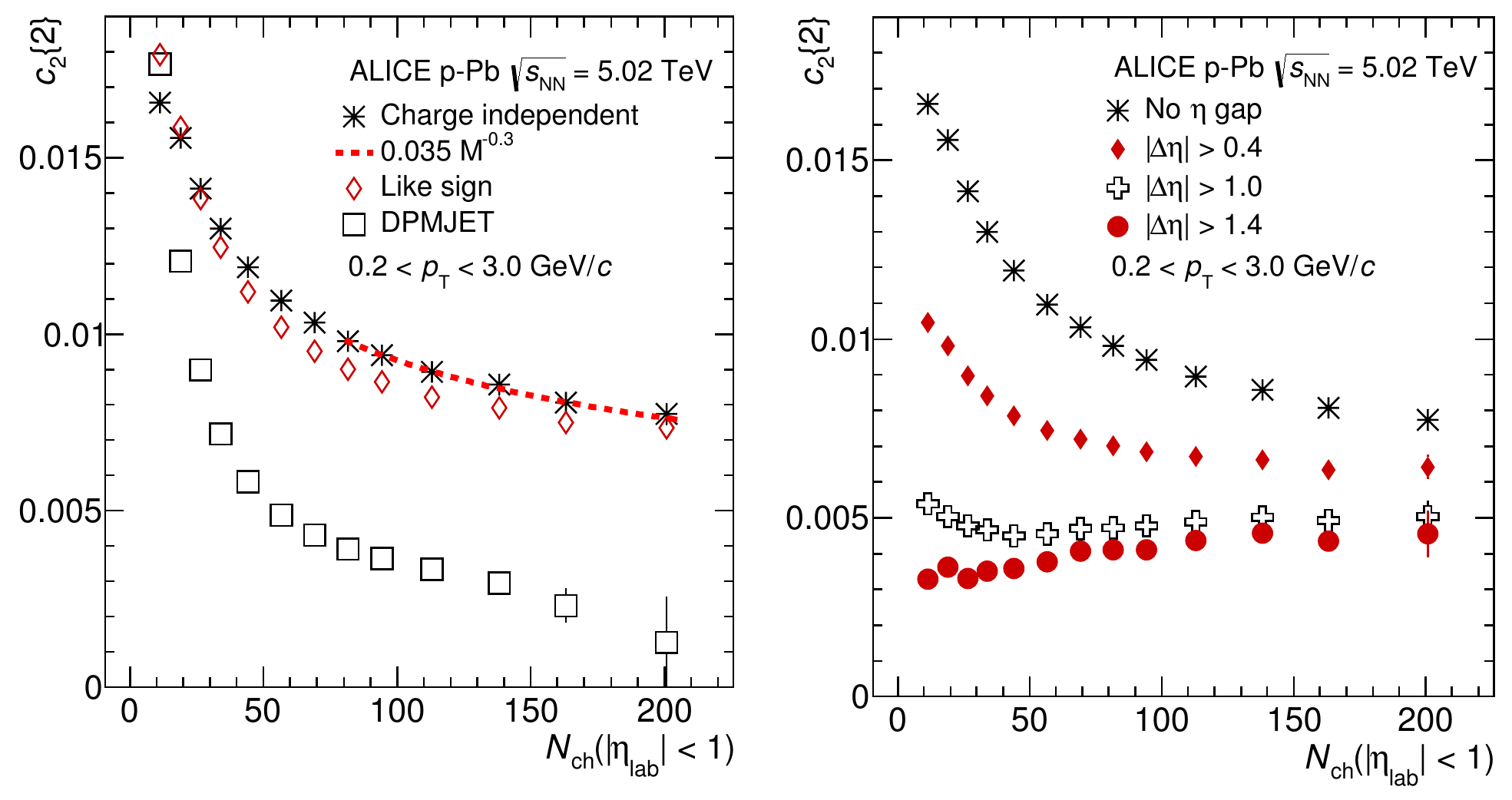}
\caption{Mid-rapidity ($|\eta| < 1$) measurements of $c_{2}\{2\}$ as a function of multiplicity for $p$-Pb collisions. Only statistical errors are shown as these dominate the uncertainty. See table \ref{tab:sysPb} for systematic uncertainties.}
\label{fig:QC2Fa}
\end{figure}
\begin{figure}[th!f]
\includegraphics[width = 1\textwidth]{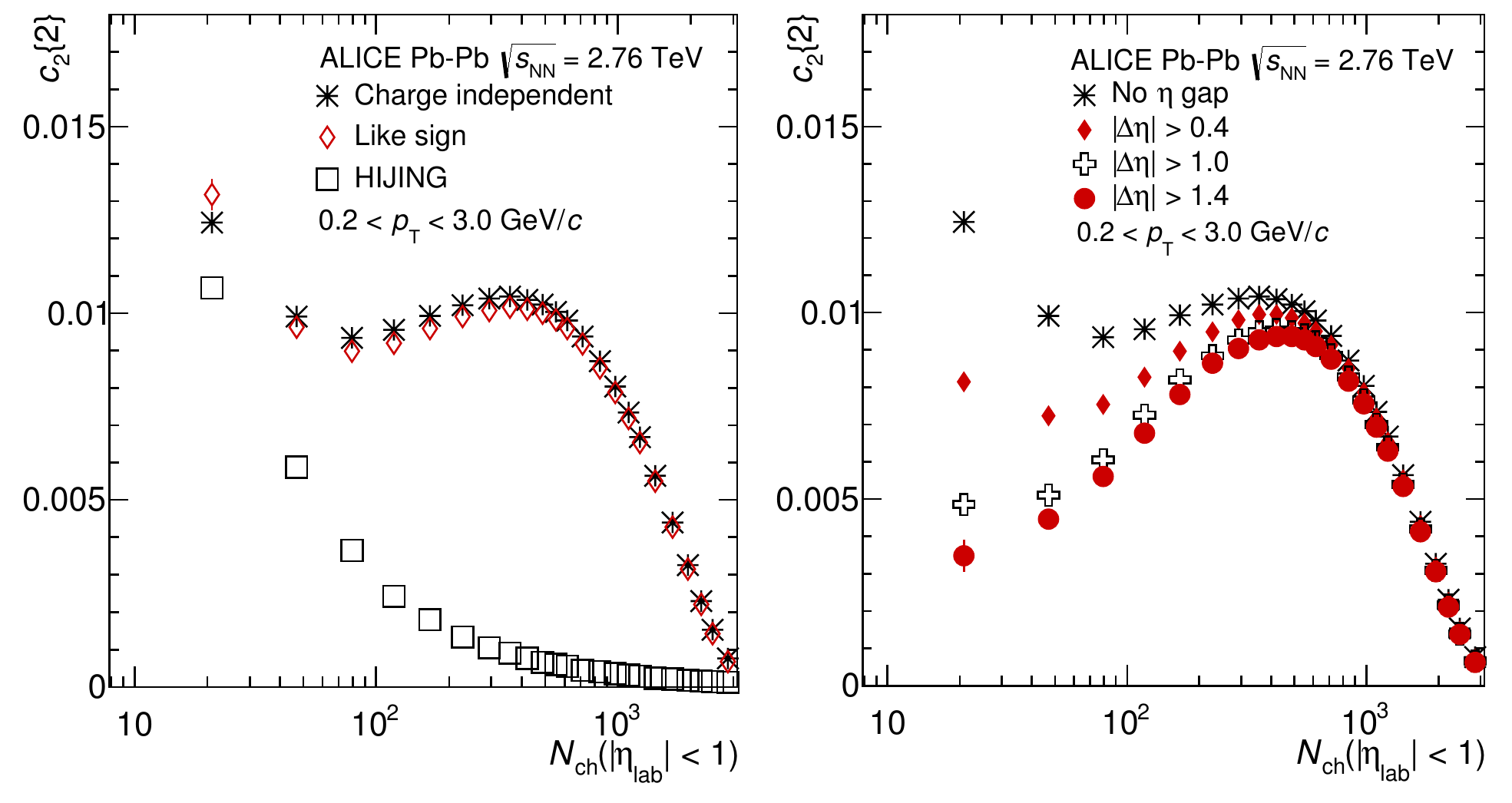}
\caption{Mid-rapidity ($|\eta| < 1$) measurements of $c_{2}\{2\}$ as a function of multiplicity in Pb-Pb collisions. Only statistical errors are shown as these dominate the uncertainty. See table \ref{tab:sysPb} for systematic uncertainties.}
\label{fig:QC2Fb}
\end{figure}
\begin{figure}[th!f]
\begin{center}
\includegraphics[width = 0.55\textwidth]{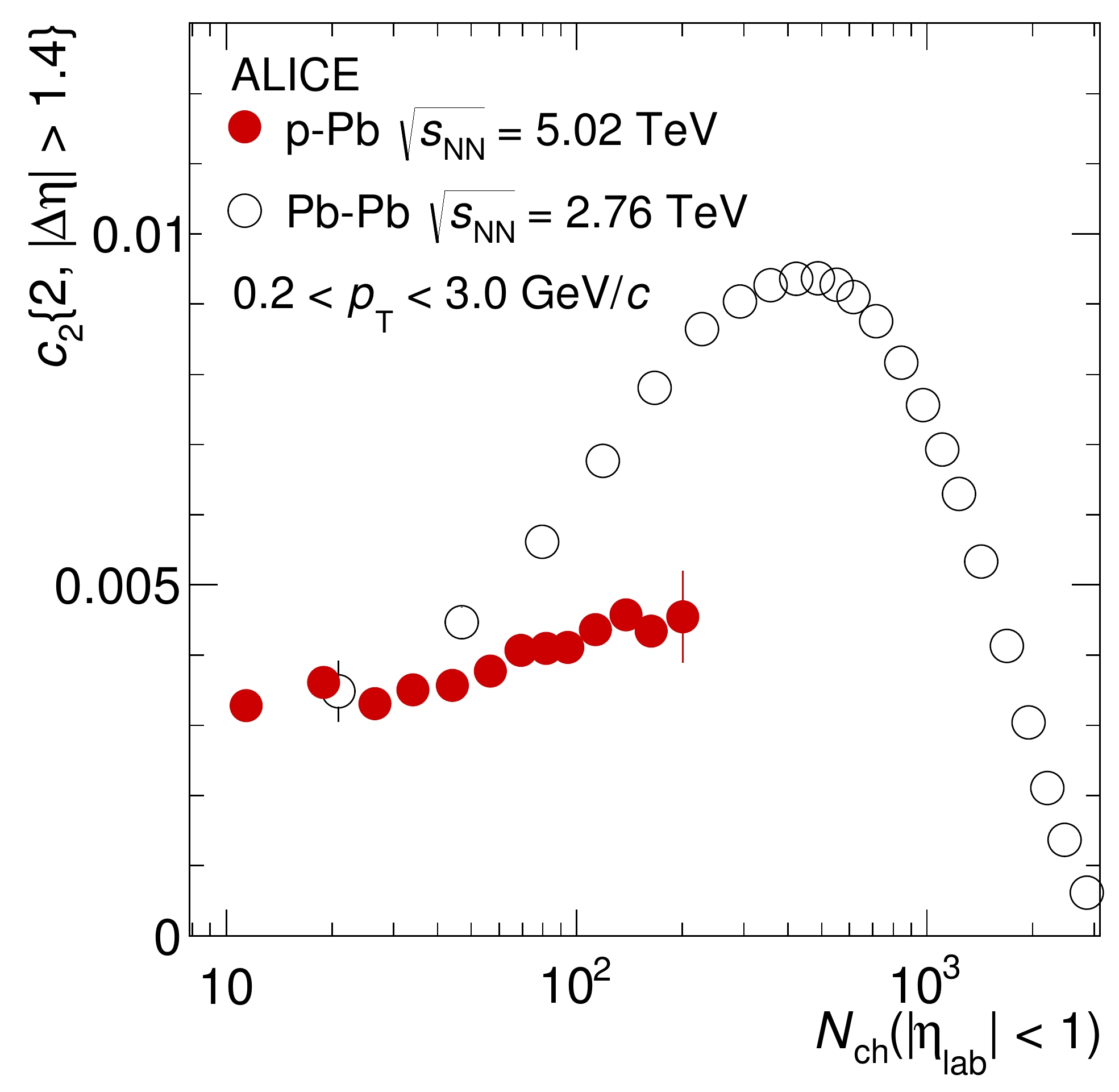}
\caption{Comparison of $c_{2}\{2\}$ with $|\Delta \eta| > 1.4 $ for $p$-Pb and Pb-Pb collisions. Only statistical errors are shown as these dominate the uncertainty. See table \ref{tab:sysPb} for systematic uncertainties.}
\label{fig:QC2c}
\end{center}
\end{figure}
The results of $c_{2}\{2\}$ as a function of multiplicity are shown in Figs. \ref{fig:QC2Fa} and \ref{fig:QC2Fb} for $p$-Pb and Pb-Pb respectively. The left column presents the results, using the $Q$-cumulants methods \cite{Bilandzic:2010jr} in the case where no $\Delta\eta$ gap is applied. Charge independent refers to the fact that all available charged tracks are used to determine the cumulants. The left panel of Fig. \ref{fig:QC2Fa} shows that the star symbols (charge independent measurements) in $p$-Pb collisions exhibit a decrease with increasing multiplicity, qualitatively consistent with the expectation of correlations dominated by non flow effects. When fitting these data points with the function $a /M^{b} $ at large multiplicity, we find $b = 0.3$. The value $b=1$ is expected if high-multiplicity events are a linear superposition of low multiplicity events \cite{Voloshin:2008dg}. This deviation from 1 might indicate the existence of another mechanism that increases $c_{2}\{2\}$, or that the relative fraction of few particle correlations is increasing with multiplicity. In the same plot, we present measurements of like--sign correlations, calculated by measuring $c_{2}\{2\}$ for positive and negative tracks separately, and forming the average. The corresponding points, represented by the diamonds, are lower than the charge independent results for the majority of the multiplicity ranges. This is expected since few-particle correlations from jets and resonances conserve charge, and thus are more likely to be absent in the like-sign measurements. Conversely, the like-sign measurements are higher for the lowest multiplicity bin. This can be explained by a suppression of unlike sign correlations (e.g. multi-particle jets) induced by the low multiplicity cut. Our results in $p$-Pb collisions are compared to predictions from the DPMJET model \cite{Roesler:2000he}. It includes in a phenomenological way the soft multi-particle production as well as hard scatterings, contains no collective effects and thus can serve as a benchmark to study the effect of non flow on our measurements. It is seen that the corresponding points for $c_{2}\{2\}$ in DPMJET fall off more rapidly compared to data. When carrying out the $a /M^{b}$ fit to the model, we find $b \sim 0.8$. The data is also significantly higher than DPMJET at high multiplicity. 
\begin{figure}[t]
\includegraphics[width = 1\textwidth]{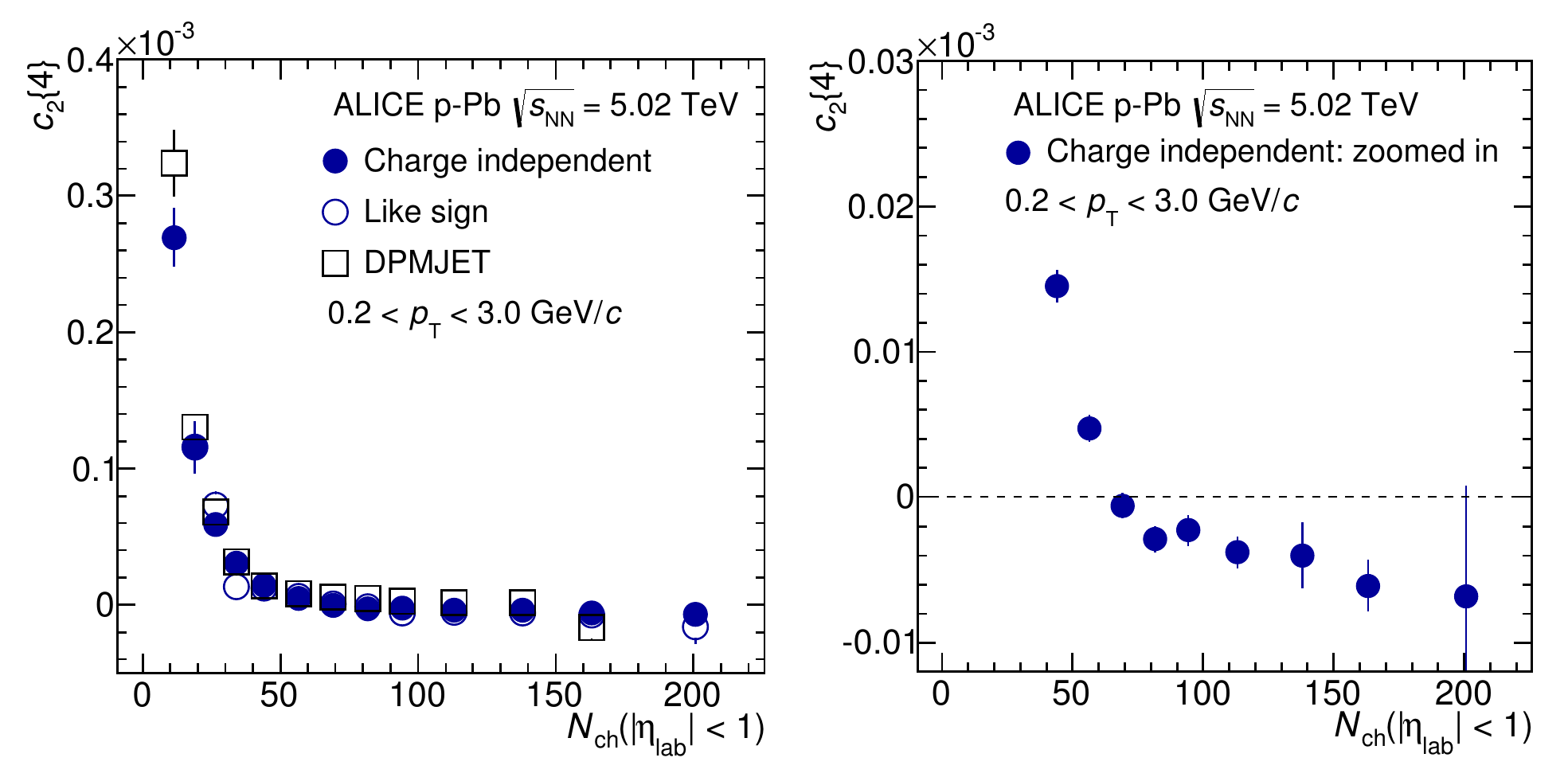}
\caption{Mid-rapidity ($|\eta| < 1$) measurements of $c_{2}\{4\}$ as a function of multiplicity for $p$-Pb collisions. Only statistical errors are shown as these dominate the uncertainty. See table \ref{tab:sysPb} for systematic uncertainties. The right panel shows a zoomed in version of the solid points in the left panel.}
\label{fig:QC4Fa}
\end{figure}
\begin{figure}[t]
\begin{tabular}{clc}
\includegraphics[width = 0.5\textwidth]{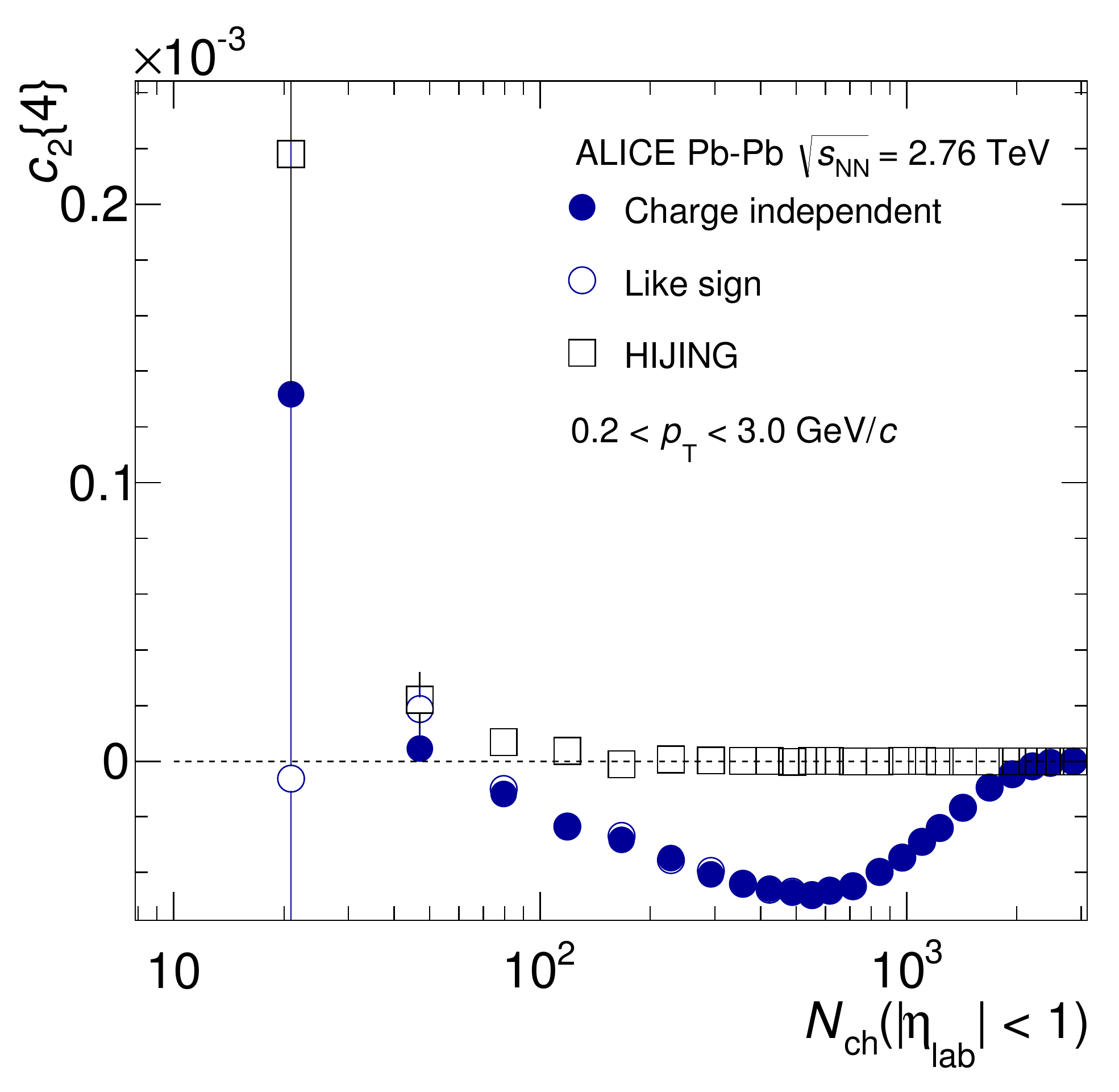}
&
\includegraphics[width = 0.5\textwidth]{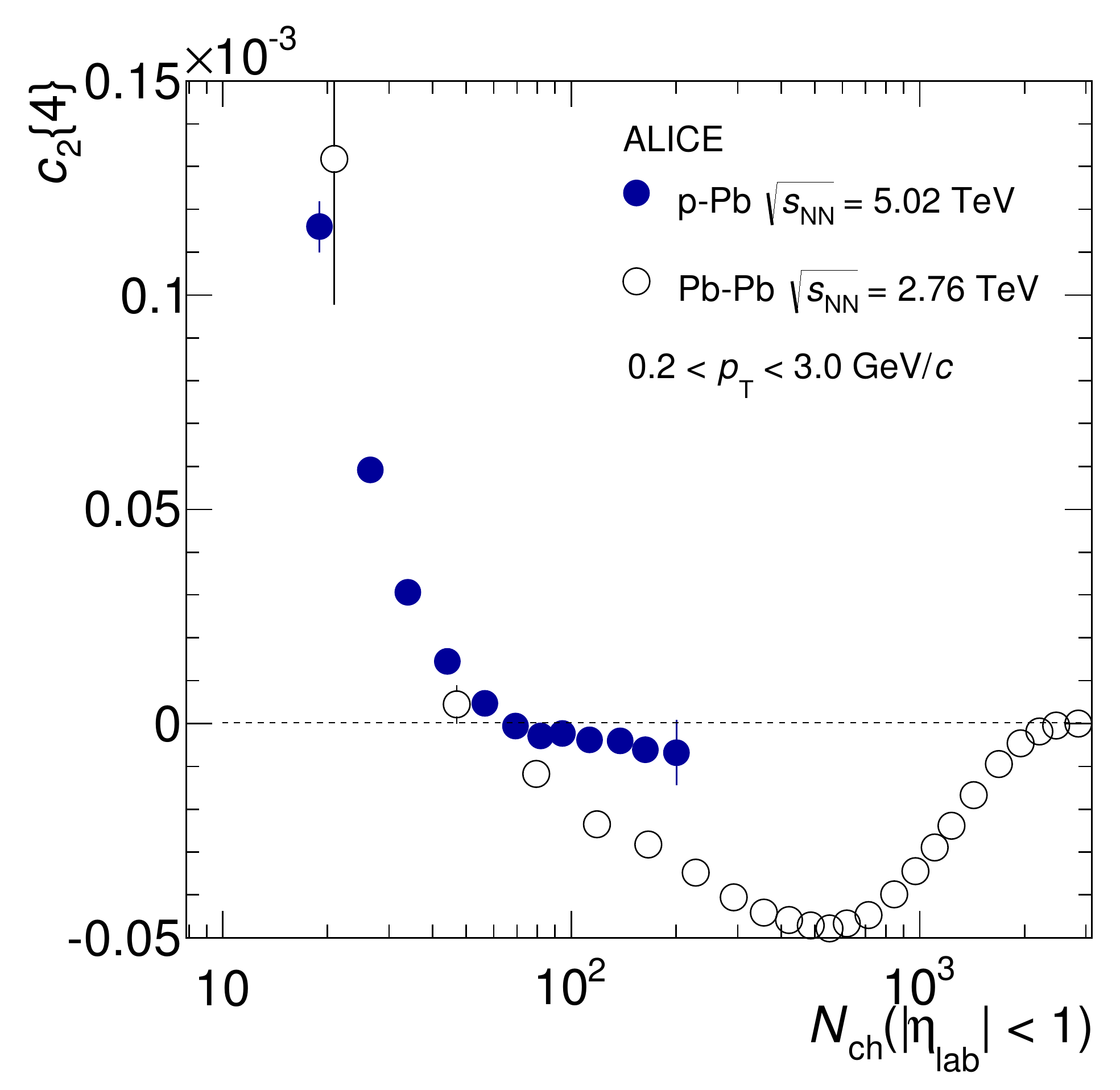}
\end{tabular}
\caption{Left Panel: Mid-rapidity ($|\eta| < 1$) measurements of $c_{2}\{4\}$ as a function of multiplicity for Pb-Pb collisions. Right Panel: Comparison of $c_{2}\{4\}$ for $p$-Pb and Pb-Pb collisions. Only statistical errors are shown as these dominate the uncertainty. See table \ref{tab:sysPb} for systematic uncertainties.}
\label{fig:QC4Fb}
\end{figure}

The right panel of Fig. \ref{fig:QC2Fa} presents the multiplicity dependence of the two-particle cumulants in $p$-Pb collisions in the case where a $\Delta\eta$ gap is applied. It is seen that for a given multiplicity, increasing the gap decreases $c_{2}\{2\}$. As mentioned previously, this is expected since tracks from few-particle correlations such as jets and resonances have smaller relative angles, therefore their contribution is suppressed by the applied pseudo-rapidity separation. However for large $\Delta\eta$ values, i.e.~for $|\Delta \eta| > 1$, the data points increase with multiplicity which is not expected if non flow dominates. In addition, the $|\Delta \eta|$ dependence of $c_{2}\{2\}$ is less pronounced at higher multiplicities. This could be a consequence of a flow-like mechanism with no or little dependence on $\eta$, whose relative strength increases with increasing multiplicity.

The Pb-Pb results of $c_{2}\{2\}$ in the case of the charge independent and the like-sign analysis are presented in the left panel of Fig. \ref{fig:QC2Fb}. They decrease with increasing multiplicity up to $N_{\rm ch}\sim100$, then increase until mid--central collisions (i.e.~up to $N_{\rm ch} \approx 400$). When moving to more central events where initial state anisotropies decrease, the values of $c_{2}\{2\}$ decrease as expected. Predictions from the HIJING model are also shown in the same plot. This model, similarly to the DPMJET model, contains only non flow, and as expected, $c_{2}\{2\}$ attenuates more rapidly than the data. Finally, the right panel of Fig. \ref{fig:QC2Fb} presents the two-particle results in Pb-Pb collisions after applying a $\Delta\eta$ gap to reduce the contribution from non flow. It is seen that at multiplicities $N_{\rm ch} \gtrsim 1000$, the measurements with various $\Delta \eta$ gaps converge, indicating the dominance of anisotropic flow. The measurements at lower multiplicities depend on $\Delta \eta$ gap significantly, indicating non flow plays a prominent role. 

In Fig. \ref{fig:QC2c}, we compare $c_{2}\{2\}$ for $p$-Pb and Pb-Pb with  $|\Delta \eta| > 1.4 $ to minimize the contribution from non flow.  Both systems have similar values of  $c_{2}\{2\}$ at low multiplicity, however the Pb-Pb data points rise more rapidly for higher multiplicities. This maybe explained by higher eccentricities (therefore higher anisotropies) in Pb-Pb collisions found from a CGC inspired cluster model for the initial conditions at similar multiplicities \cite{Basar:2013hea} (not shown). We note that other studies are exploring these correlations with the AMPT model \cite{Ma:2014pva}.
\begin{figure}[t]
\begin{center}
\includegraphics[width = 0.55\textwidth]{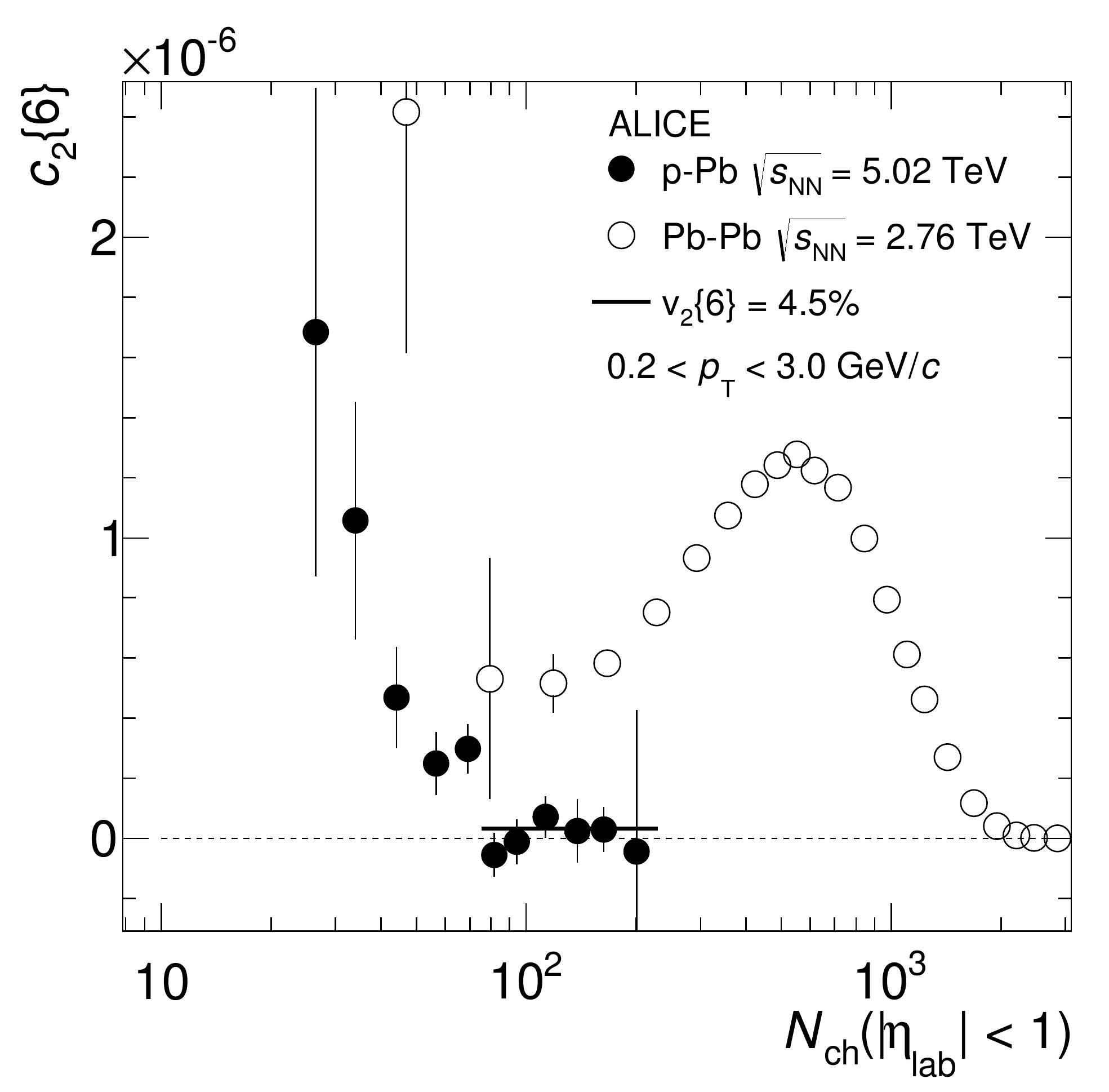}
\caption{Comparison of mid-rapidity ($|\eta| < 1$) $c_{2}\{6\}$ for $p$-Pb and Pb-Pb collisions. Only statistical errors are shown as these dominate the uncertainty. See table \ref{tab:sysPb} for systematic uncertainties.}
\label{fig:QC6c}
\end{center}
\end{figure}
\begin{figure}[tbh!f]
\begin{center}
\includegraphics[width = 0.55\textwidth]{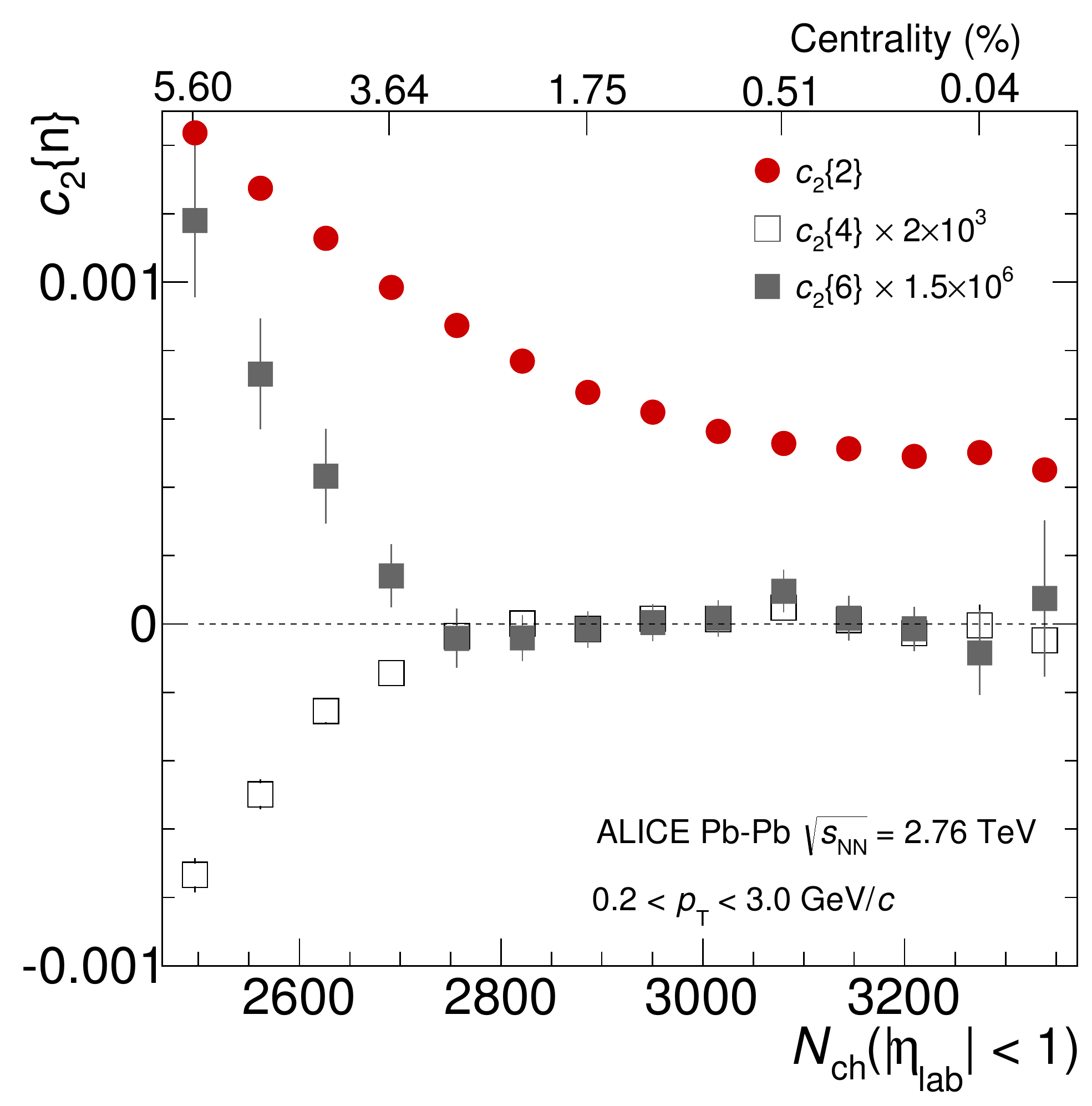}
\caption{Comparison of $c_{2}\{m\}$ in very high-multiplicity Pb-Pb collisions. Only statistical errors are shown as these dominate the uncertainty. See table \ref{tab:sysPb} for systematic uncertainties.}
\label{fig:QCVC}
\end{center}
\end{figure}
\begin{figure}[tbh!f]
\includegraphics[width = 1\textwidth]{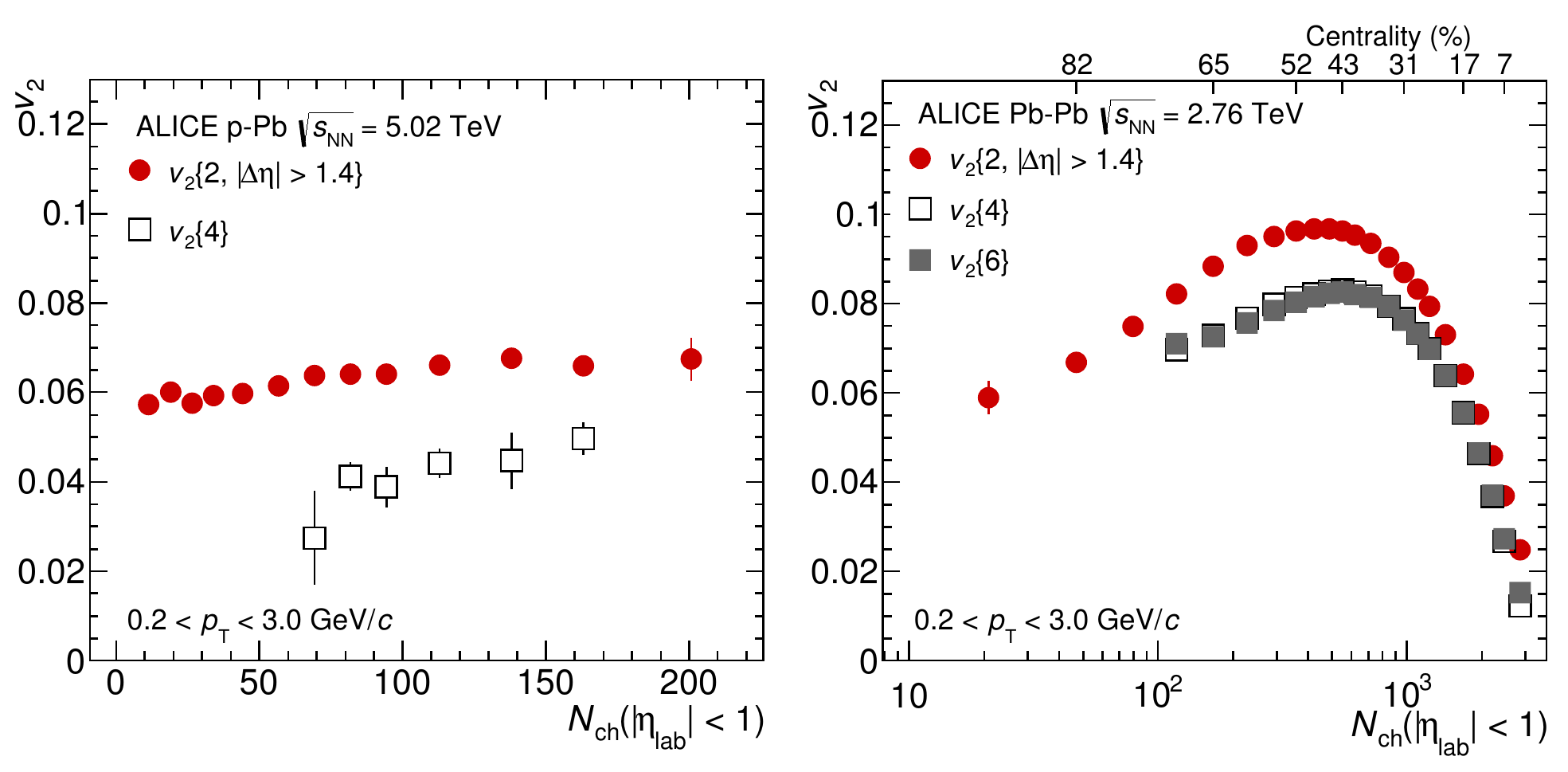}
\caption{Measurements of $v_{2}\{2\}$, $v_{2}\{4\}$, and $v_{2}\{6\}$ in $p$-Pb (left panel) and Pb-Pb (right panel) collisions. The measurements of $v_{2}\{2\}$ are obtained with a $|\Delta \eta| > 1.4$ gap. Only statistical errors are shown as these dominate the uncertainty. See table \ref{tab:sysPb} for systematic uncertainties.}
\label{fig:v_nsa}
\end{figure}
\begin{figure}[tbh!f]
\includegraphics[width = 1\textwidth]{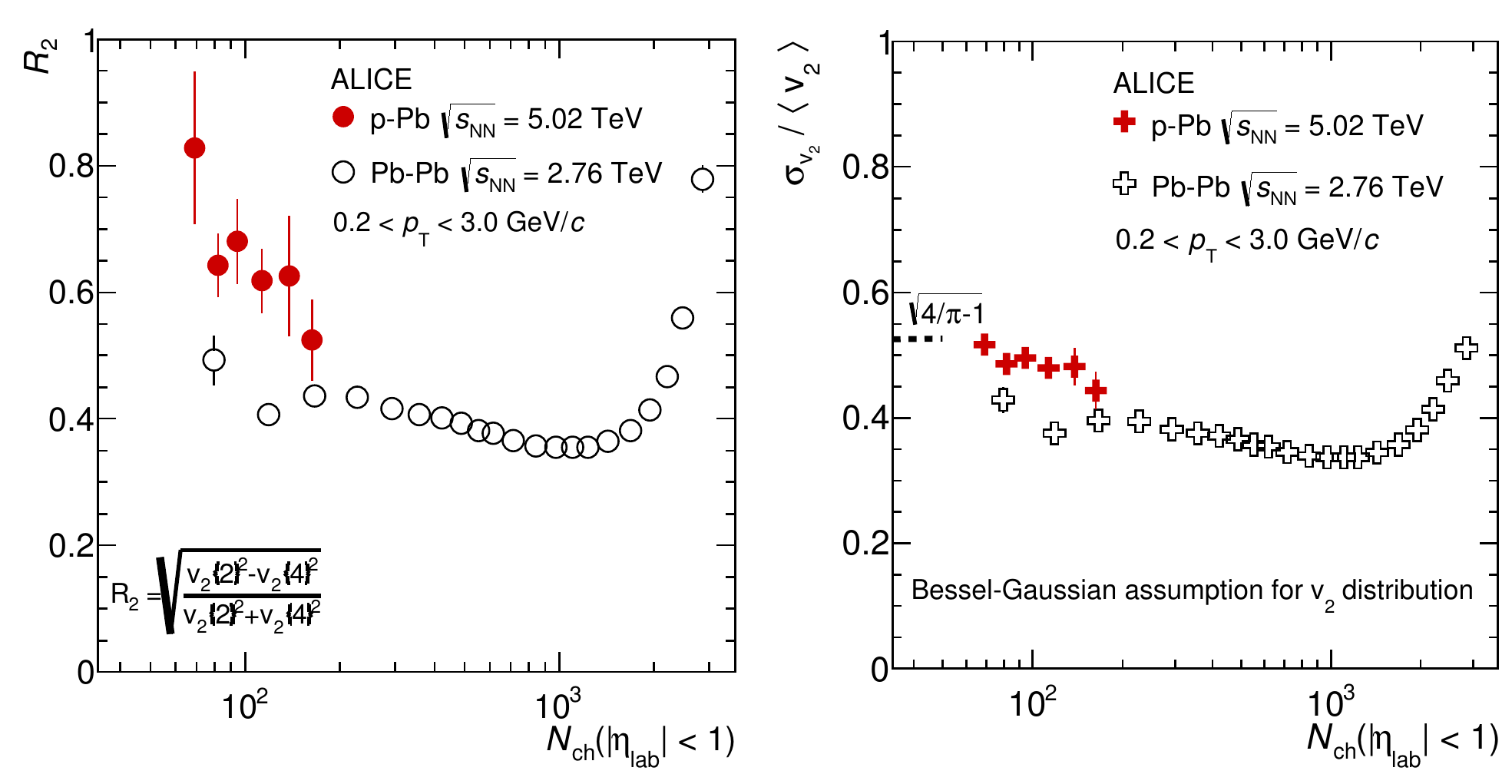}
\caption{Left Panel: Measurements of $[ (v_{2}\{2\}^{2}-v_{2}\{4\}^{2})/(v_{2}\{2\}^{2}+v_{2}\{4\}^{2})]^{1/2}$ in $p$-Pb and Pb-Pb collisions. The measurements of $v_{2}\{2\}$ are obtained with a $|\Delta \eta| > 1.4$ gap. Only statistical errors are shown as these dominate the uncertainty. See table \ref{tab:sysPb} for systematic uncertainties. Right Panel: $\sigma_{v2} / \langle v_{2} \rangle$ obtained from the same $v_{2}\{2\}$ and $v_{2}\{4\}$ measurements assuming a Bessel-Gaussian distribution.}
\label{fig:v_nsb}
\end{figure}

\subsection{The second harmonic four-particle cumulant}

The results of $c_{2}\{4\}$ as a function of multiplicity are shown in Fig.~\ref{fig:QC4Fa} for $p$-Pb collisions, and Fig.~\ref{fig:QC4Fb} for Pb-Pb collisions. We use the $Q$-cumulants methods to obtain the results in all cases. For $p$-Pb collisions, there are little differences between the like-sign and the charge independent results. The values of $c_{2}\{4\}$  attenuate more rapidly than $c_{2}\{2\}$ at low multiplicity, as expected since non flow contributes significantly in this region. The predictions from the DPMJET model, represented by the open squares in Fig.~\ref{fig:QC4Fa}, also show a large attenuation. At $N_{\rm ch} \gtrsim 70$, the values of $c_{2}\{4\}$ become negative, and this is illustrated in the right panel of Fig.~\ref{fig:QC4Fa}. Measurements of $c_{2}\{4\}$ below zero allow for real values of $v_{2}\{4\}$. We found that the position of the transition from positive to negative depends on the $\eta$ cut applied to the tracks (not shown). When the $\eta$ cut is reduced, the transition occurs at a larger multiplicity, which is presumably due to the larger contribution of non flow. The results for Pb-Pb collisions shown in the left panel of Fig.~\ref{fig:QC4Fb} with the circles, exhibit a similar trend. The values of $c_{2}\{4\}$ rise at very high multiplicities as the collisions become central. The charge independent HIJING predictions, also shown in this plot as open squares, converge to zero for most multiplicities indicating the contribution from non flow is negligible. In the right panel of Fig.~\ref{fig:QC4Fb}, we compare $c_{2}\{4\}$ for $p$-Pb and Pb-Pb collisions. Both systems exhibit positive values for $N_{\rm ch} \lesssim 70$, indicating a dominance of non flow. At multiplicities $70 \lesssim N_{\rm ch} \lesssim 200$, $c_{2}\{4\}$ decreases more rapidly for Pb-Pb which might be indicative of higher eccentricities for similar multiplicities.

\subsection{The second harmonic six-particle cumulant}

The results of $c_{2}\{6\}$ as a function of multiplicity are shown in Fig.~\ref{fig:QC6c} for $p$-Pb and Pb-Pb collisions. We again use the $Q$-cumulants methods to obtain $c_{2}\{6\}$. In $p$-Pb collisions, these measurements are more limited by finite statistics as we observe fluctuations above and below zero at high multiplicity (within the statistical uncertainties). The solid black line indicates $v_{2}\{6\}=4.5 \%$, which is roughly the value of $v_{2}\{4\}$ in this multiplicity region. The $p$-Pb measurements will benefit from higher statistics measurements planned for future LHC running. However, it is clear at multiplicities above 100, the values of $c_{2}\{6\}$ are significantly higher for Pb-Pb compared to $p$-Pb. This again maybe be explained by higher eccentricities in the initial state of the colliding nuclei for the former.

\begin{figure}[t]
\includegraphics[width = 1\textwidth]{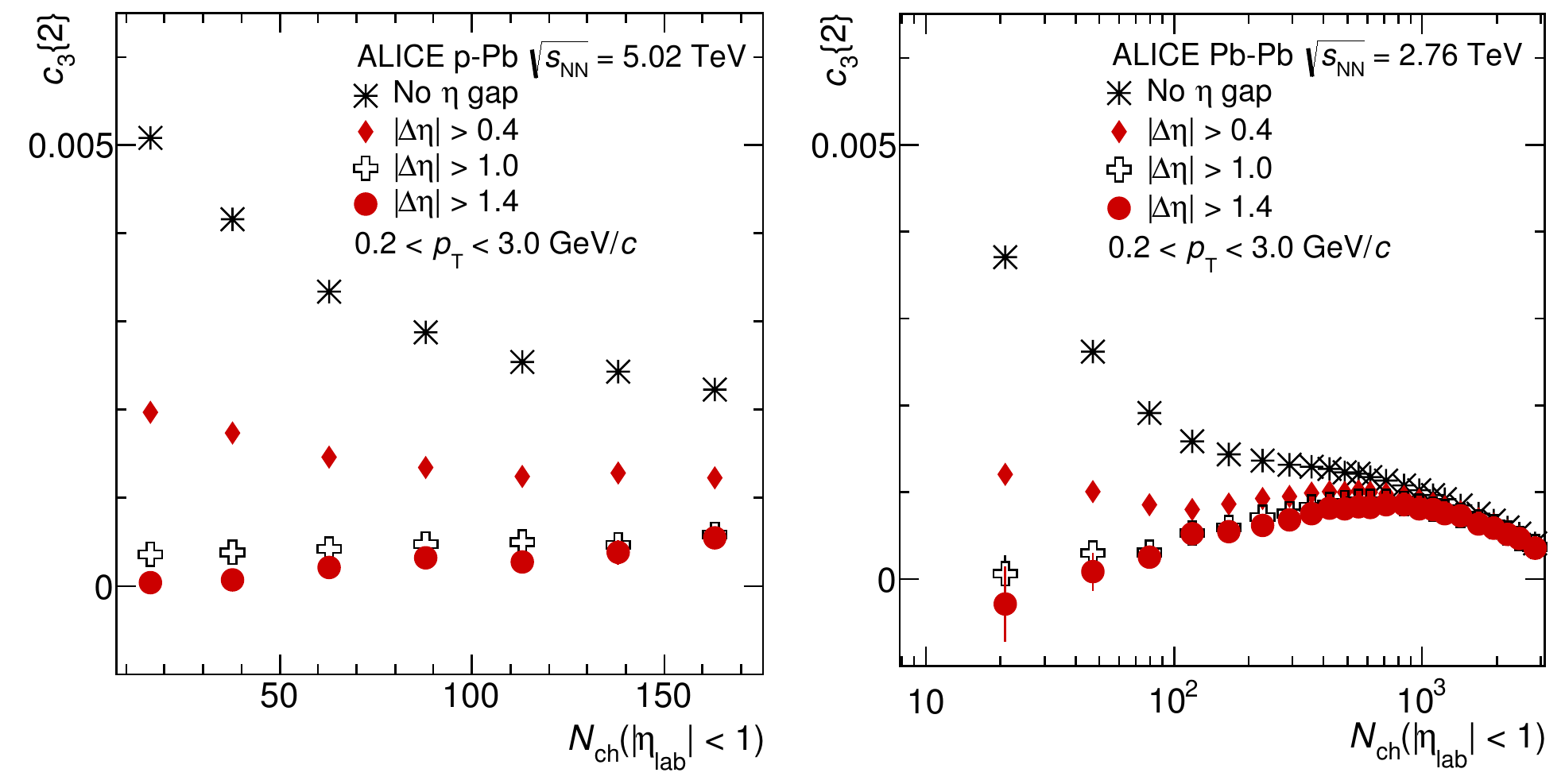}
\caption{Third harmonic two-particle cumulants in $p$-Pb and Pb-Pb collisions. Only statistical errors are shown as these dominate the uncertainty. See table \ref{tab:sysPb} for systematic uncertainties.}
\label{fig:v3a}
\end{figure}

\begin{figure}[t]
\begin{center}
\includegraphics[width = 0.55\textwidth]{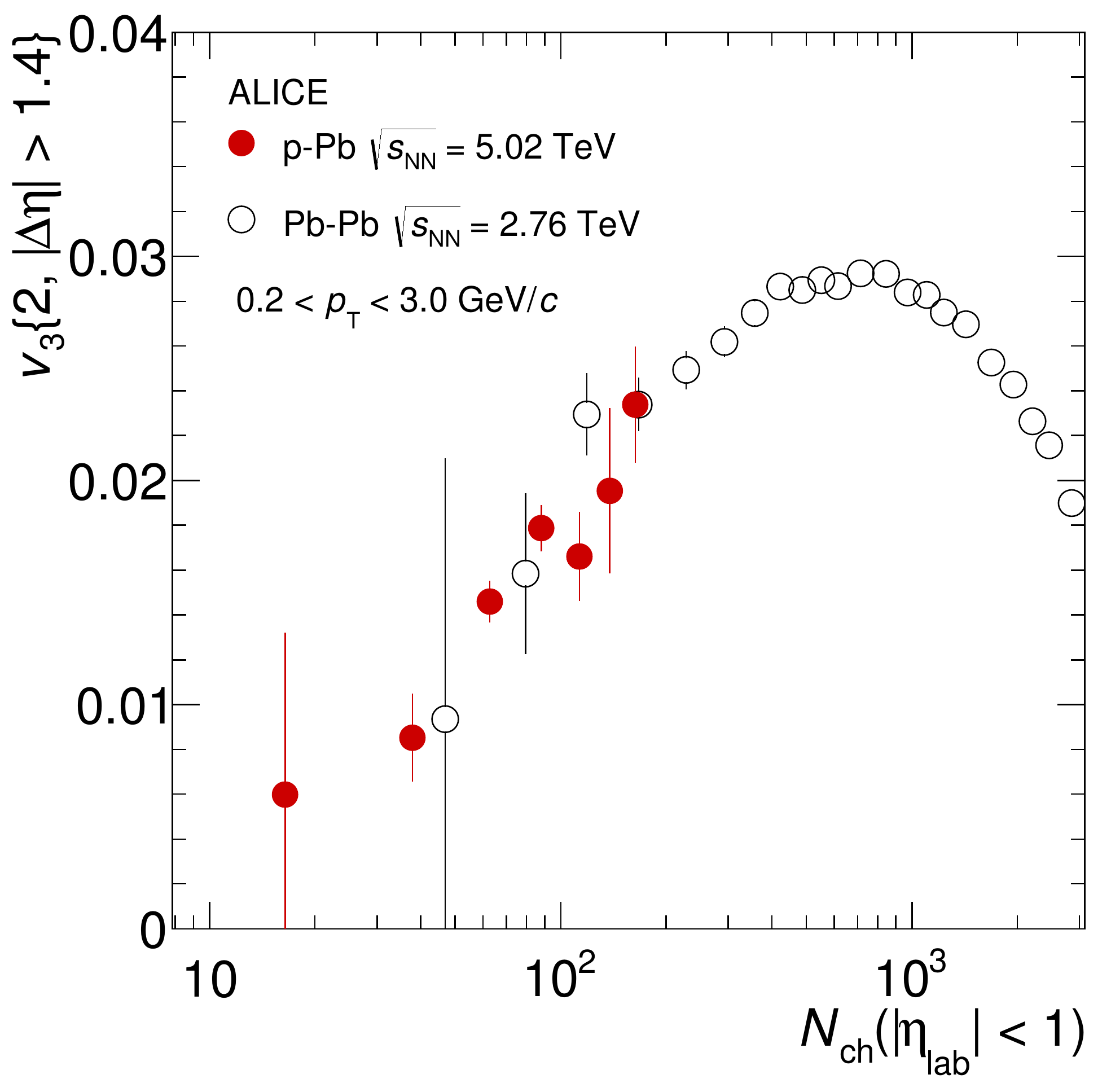}
\caption{Third harmonic flow coefficients in $p$-Pb and Pb-Pb collisions. The measurements of $v_{3}\{2\}$ are obtained with a $|\Delta \eta| > 1.4$ gap.  Only statistical errors are shown as these dominate the uncertainty. See table \ref{tab:sysPb} for systematic uncertainties.}
\label{fig:v3b}
\end{center}
\end{figure}

\subsection{Second harmonic cumulants in very high-multiplicity Pb-Pb collisions}

The non-zero values of $c_{2}\{4\}$ in high-multiplicity $p$-Pb collisions merit a comparison to high-multiplicity Pb-Pb collisions, which have an impact parameter that becomes small. In both cases, initial state fluctuations are expected to dominate the eccentricity since there is no intrinsic eccentricity from the overlapping nuclei. In Fig.~\ref{fig:QCVC}, cumulants of different orders are compared for high-multiplicity Pb-Pb collisions. At $N_{\rm ch} \gtrsim 2800$, $c_{2}\{4\}$ becomes consistent with zero, which is in contrast to high-multiplicity $p$-Pb (where $c_{2}\{4\}$ is negative). The measurements of $c_{2}\{6\}$ also become zero in exactly the same region, which corresponds to the highest $\sim 2.5\%$ of the cross section. Constant fits to $c_{2}\{4\}$ and  $c_{2}\{6\}$ for $N_{\rm ch} > 2800$ give $8.5 \times 10^{-6} \pm 9.3 \times 10^{-6}$ and $7.2 \times 10^{-6} \pm 2.2 \times 10^{-5}$ respectively (with $\chi^2/dof \sim 1$ in each case). An explanation for the difference between $p$-Pb and Pb-Pb can be found by considering the number of sources which form the eccentricity. When this number is small, eccentricity fluctuations have a power-law distribution which will lead to finite values of $c_{2}\{4\}$ and $c_{2}\{6\}$, assuming $v_{2} \propto \varepsilon_{2}$ \cite{Yan:2013laa}. When the number of sources becomes large enough, the power-law distribution becomes equivalent to the Bessel-Gaussian distribution \cite{Voloshin:2007pc, Bzdak:2013rya}. In the special case of very high multiplicity Pb-Pb collisions where the impact parameter is expected to approach 0, the Bessel-Gaussian distribution gives values of $c_{2}\{4\}$ and $c_{2}\{6\}$ that are zero. Assuming the number of sources are highly correlated with the number of participants, the difference between very high multiplicity $p$-Pb and Pb-Pb can be explained by the larger number of sources in the latter. Finally, these results at the LHC can be compared to those from the STAR Collaboration \cite{Agakishiev:2011eq, Wang:2014}. In Au-Au $\sqrt{s_{\rm NN}} = 200$~GeV collisions, $c_{2}\{4\}$ also approaches zero and may become  positive which prevented the extraction of $v_{2}\{4\}$ in central collisions, while for U-U $\sqrt{s_{\rm NN}} = 193$~GeV collisions, $c_{2}\{4\}$ always remains negative.

\subsection{Second harmonic flow coefficients in $p$-Pb and Pb-Pb collisions}

A comparison of second harmonic flow coefficients is shown in Fig.~\ref{fig:v_nsa}. We determine $v_{2}\{2\}$ with the largest possible $\Delta \eta$ gap to minimize the contribution from non flow. In $p$-Pb collisions, we find $v_{2}\{2\} > v_{2}\{4\}$ which is indicative of flow fluctuations, but can also be affected by non flow. The same observation is made for Pb-Pb collisions, and we also find $v_{2}\{4\} \simeq v_{2}\{6\}$. Regarding the functional form of the $v_{2}$ distribution, a Bessel-Gaussian function satisfies the criterium $v_{2}\{4\}=v_{2}\{6\}$ \cite{Voloshin:2007pc}. When the Bessel function of the Bessel-Gaussian becomes 1, $v_{2}\{4\}=v_{2}\{6\}=0$. A power-law function gives values of $v_{2}\{4\}$ and $v_{2}\{6\}$ which are close, but not exactly equal \cite{Yan:2013laa}. In addition, unfolded measurements of the $v_2$ distribution have shown Bessel-Gaussian descriptions work reasonably well for Pb-Pb collisions \cite{Timmins:2013hq, Aad:2013xma}. In the left panel of Fig.~\ref{fig:v_nsb}, we show the measurement of $R_{2}$, defined as:
\begin{equation} 
\label{eq:R}
R_{n} =  \sqrt{\frac{v_{n}\{2\}^{2}-v_{n}\{4\}^{2}} {v_{n}\{2\}^{2}+v_{n}\{4\}^{2}}}
\end{equation} 
As mentioned in Section II, when $\sigma_{v_n} \ll \langle v_n \rangle$, $R_{n} = \sigma_{v_n}/ \langle v_{n} \rangle$ in case non flow is negligible. In the overlapping multiplicities, the values for $p$-Pb appear to be higher than Pb-Pb, demonstrating the greater role of fluctuations in the former. A similar observation is reported by the CMS Collaboration \cite{Chatrchyan:2013nka}. The trend for $R_{2}$ in Pb-Pb is similar to observations for Au-Au $\sqrt{s_{\rm NN}} = 200$~GeV collisions \cite{Agakishiev:2011eq, Alver:2010rt}. The value of $R_{2}$ in mid-central (mid-multiplicity) Pb-Pb collisions ($\sim0.35$) is between the STAR and PHOBOS results for similar centralities. In the right panel, we show $\sigma_{v2} /\langle v_{2} \rangle$ under the assumption that the $v_2$ distribution is Bessel-Gaussian. Using this assumption, all the information from distribution can be obtained from just $v_{2}\{2\}$ and $v_{2}\{4\}$, without the need for the condition $\sigma_{v_n} \ll v_n$ \cite{Voloshin:2007pc}. The dashed lines denote the $\sigma_{v2} /\langle v_{2} \rangle=\sqrt{4/\pi-1}$ limit, expected when fluctuations dominated the eccentricity \cite{Broniowski:2007ft}. We find that the Bessel-Gaussian $\sigma_{v2} /\langle v_{2} \rangle$ is close to this limit for high-multiplicity Pb-Pb collisions.

\subsection{Two-particle cumulants of the third harmonic}

In Fig.~\ref{fig:v3a}, we show measurements of the third harmonic two-particle cumulants for $p$-Pb and Pb-Pb collisions, for different values of the $\Delta \eta$ gap. For $p$-Pb and low Pb-Pb multiplicities, we generally find a strong dependence on the $\Delta \eta$. The values with small $\Delta \eta$ gaps decrease with multiplicity in $p$-Pb, as expected when non flow is dominant. This behavior was also observed by the STAR Collaboration at lower beam energies \cite{Adamczyk:2013waa}. The measurements with larger $\Delta \eta$ gaps show an increase with multiplicity, indicating a contribution from global correlations. For large Pb-Pb multiplicities, measurements with various $\Delta \eta$ gaps converge indicating a dominance of flow. Finally, in Fig.~\ref{fig:v3b} we compare the third harmonic flow coefficients for both systems, again with the largest possible $\Delta \eta$ gap. In contrast to measurements of the second harmonic, we find that $p$-Pb and Pb-Pb are consistent for the same multiplicity. This consistency has also been observed by the CMS Collaboration \cite{Chatrchyan:2013nka}, and points to similar third harmonic eccentricities for $p$-Pb and Pb-Pb at the same multiplicity. A CGC inspired cluster model for the initial conditions is able to reproduce this observation \cite{Basar:2013hea}.

\section{Summary}
We have reported results of $c_{2}\{2\}$, $c_{2}\{4\}$, and $c_{2}\{6\}$ as a function of multiplicity in \pPb and \Pb collisions for kinematic cuts $0.2 < p_T < 3$ GeV/$c$ and $|\eta| < 1$. Measurements of $c_{2}\{2\}$ using all pairs in the event for $p$-Pb collisions show a decrease with multiplicity, characteristic of a dominance of few-particle correlations. However, the decrease is shallower than from the expectation high-multiplicity events are a superposition of low multiplicity events. When a $|\Delta \eta|$ gap is placed to suppress such non flow correlations, measurements of $c_{2}\{2\}$ begin to rise at high-multiplicity. Similar observations are made for Pb-Pb collisions. The measurements of $c_{2}\{4\}$ exhibit a transition from positive values at low multiplicity to negative values at higher multiplicity for both $p$-Pb and Pb-Pb. The negative values allow for a real $v_{2}\{4\}$, which is lower than $v_{2}\{2\}$ at a given multiplicity. The measurements of $c_{2}\{6\}$ for $p$-Pb collisions are both consistent with zero, and the assumption $v_{2}\{4\}=v_{2}\{6\}$. In Pb-Pb collisions, we observe $v_{2}\{4\} \simeq v_{2}\{6\}$, which is indicative of a Bessel-Gaussian function for the $v_{2}$ distribution in this domain. For very high-multiplicity Pb-Pb collisions, both $v_{2}\{4\}$ and $v_{2}\{6\}$ are consistent with 0. A comparison of $p$-Pb cumulants to those of Pb-Pb at the same multiplicity (for $N_{ch} \gtrsim 70$) shows stronger correlations in Pb-Pb for all the cumulants. This may be explained by higher eccentricities for similar multiplicities. Finally, we have performed measurements of $v_{3}\{2\}$ for $p$-Pb and Pb-Pb collisions. They are found to be similar for overlapping multiplicities when a $|\Delta \eta| > 1.4$ gap is placed, indicating initial state third harmonic eccentricities may be similar for both systems. We conclude that our measurements indicate the (double) ridge observed in \pPb arises from global azimuthal correlations, rather than from few-particle correlations which decrease with multiplicity. These measurements provide key constraints to the initial state and transport properties in $p$-Pb and Pb-Pb collisions.

%
%

\newenvironment{acknowledgement}{\relax}{\relax}
\begin{acknowledgement}
\section*{Acknowledgements}
The ALICE Collaboration would like to thank all its engineers and technicians for their invaluable contributions to the construction of the experiment and the CERN accelerator teams for the outstanding performance of the LHC complex.
The ALICE Collaboration gratefully acknowledges the resources and support provided by all Grid centres and the Worldwide LHC Computing Grid (WLCG) collaboration.
The ALICE Collaboration acknowledges the following funding agencies for their support in building and
running the ALICE detector:
State Committee of Science,  World Federation of Scientists (WFS)
and Swiss Fonds Kidagan, Armenia,
Conselho Nacional de Desenvolvimento Cient\'{\i}fico e Tecnol\'{o}gico (CNPq), Financiadora de Estudos e Projetos (FINEP),
Funda\c{c}\~{a}o de Amparo \`{a} Pesquisa do Estado de S\~{a}o Paulo (FAPESP);
National Natural Science Foundation of China (NSFC), the Chinese Ministry of Education (CMOE)
and the Ministry of Science and Technology of China (MSTC);
Ministry of Education and Youth of the Czech Republic;
Danish Natural Science Research Council, the Carlsberg Foundation and the Danish National Research Foundation;
The European Research Council under the European Community's Seventh Framework Programme;
Helsinki Institute of Physics and the Academy of Finland;
French CNRS-IN2P3, the `Region Pays de Loire', `Region Alsace', `Region Auvergne' and CEA, France;
German BMBF and the Helmholtz Association;
General Secretariat for Research and Technology, Ministry of
Development, Greece;
Hungarian OTKA and National Office for Research and Technology (NKTH);
Department of Atomic Energy and Department of Science and Technology of the Government of India;
Istituto Nazionale di Fisica Nucleare (INFN) and Centro Fermi -
Museo Storico della Fisica e Centro Studi e Ricerche "Enrico
Fermi", Italy;
MEXT Grant-in-Aid for Specially Promoted Research, Ja\-pan;
Joint Institute for Nuclear Research, Dubna;
National Research Foundation of Korea (NRF);
CONACYT, DGAPA, M\'{e}xico, ALFA-EC and the EPLANET Program
(European Particle Physics Latin American Network)
Stichting voor Fundamenteel Onderzoek der Materie (FOM) and the Nederlandse Organisatie voor Wetenschappelijk Onderzoek (NWO), Netherlands;
Research Council of Norway (NFR);
Polish Ministry of Science and Higher Education;
National Science Centre, Poland;
Ministry of National Education/Institute for Atomic Physics and CNCS-UEFISCDI - Romania;
Ministry of Education and Science of Russian Federation, Russian
Academy of Sciences, Russian Federal Agency of Atomic Energy,
Russian Federal Agency for Science and Innovations and The Russian
Foundation for Basic Research;
Ministry of Education of Slovakia;
Department of Science and Technology, South Africa;
CIEMAT, EELA, Ministerio de Econom\'{i}a y Competitividad (MINECO) of Spain, Xunta de Galicia (Conseller\'{\i}a de Educaci\'{o}n),
CEA\-DEN, Cubaenerg\'{\i}a, Cuba, and IAEA (International Atomic Energy Agency);
Swedish Research Council (VR) and Knut $\&$ Alice Wallenberg
Foundation (KAW);
Ukraine Ministry of Education and Science;
United Kingdom Science and Technology Facilities Council (STFC);
The United States Department of Energy, the United States National
Science Foundation, the State of Texas, and the State of Ohio;
Ministry of Science, Education and Sports of Croatia and  Unity through Knowledge Fund, Croatia.
\end{acknowledgement}

\bibliographystyle{utphys}   
\bibliography{cumuRefs}

\providecommand{\href}[2]{#2}\begingroup\raggedright\begin{thebibliography}{10}

\bibitem{Voloshin:1994mz}
S.~Voloshin and Y.~Zhang, ``{Flow study in relativistic nuclear collisions by
  Fourier expansion of Azimuthal particle distributions},''
  \href{http://dx.doi.org/10.1007/s002880050141}{{\em Z.Phys.} {\bfseries C70}
  (1996) 665--672},
\href{http://arxiv.org/abs/hep-ph/9407282}{{\ttfamily arXiv:hep-ph/9407282
  [hep-ph]}}.

\bibitem{Back:2004je}
B.~Back, M.~Baker, M.~Ballintijn, D.~Barton, B.~Becker, {\em et~al.}, ``{The
  PHOBOS perspective on discoveries at RHIC},''
  \href{http://dx.doi.org/10.1016/j.nuclphysa.2005.03.084}{{\em Nucl.Phys.}
  {\bfseries A757} (2005) 28--101},
\href{http://arxiv.org/abs/nucl-ex/0410022}{{\ttfamily arXiv:nucl-ex/0410022
  [nucl-ex]}}.

\bibitem{Arsene:2004fa}
{\bfseries BRAHMS} Collaboration, I.~Arsene {\em et~al.}, ``{Quark gluon plasma
  and color glass condensate at RHIC? The Perspective from the BRAHMS
  experiment},'' \href{http://dx.doi.org/10.1016/j.nuclphysa.2005.02.130}{{\em
  Nucl.Phys.} {\bfseries A757} (2005) 1--27},
\href{http://arxiv.org/abs/nucl-ex/0410020}{{\ttfamily arXiv:nucl-ex/0410020
  [nucl-ex]}}.

\bibitem{Adcox:2004mh}
{\bfseries PHENIX} Collaboration, K.~Adcox {\em et~al.}, ``{Formation of dense
  partonic matter in relativistic nucleus-nucleus collisions at RHIC:
  Experimental evaluation by the PHENIX collaboration},''
  \href{http://dx.doi.org/10.1016/j.nuclphysa.2005.03.086}{{\em Nucl.Phys.}
  {\bfseries A757} (2005) 184--283},
\href{http://arxiv.org/abs/nucl-ex/0410003}{{\ttfamily arXiv:nucl-ex/0410003
  [nucl-ex]}}.

\bibitem{Adams:2005dq}
{\bfseries STAR} Collaboration, J.~Adams {\em et~al.}, ``{Experimental and
  theoretical challenges in the search for the quark gluon plasma: The STAR's
  critical assessment of the evidence from RHIC collisions},''
  \href{http://dx.doi.org/10.1016/j.nuclphysa.2005.03.085}{{\em Nucl.Phys.}
  {\bfseries A757} (2005) 102--183},
\href{http://arxiv.org/abs/nucl-ex/0501009}{{\ttfamily arXiv:nucl-ex/0501009
  [nucl-ex]}}.

\bibitem{Aamodt:2010pa}
{\bfseries ALICE} Collaboration, K.~Aamodt {\em et~al.}, ``{Elliptic flow of
  charged particles in Pb-Pb collisions at 2.76 TeV},''
  \href{http://dx.doi.org/10.1103/PhysRevLett.105.252302}{{\em Phys.Rev.Lett.}
  {\bfseries 105} (2010) 252302},
\href{http://arxiv.org/abs/arXiv:1011.3914}{{\ttfamily arXiv:arXiv:1011.3914
  [nucl-ex]}}.

\bibitem{ALICE:2011ab}
{\bfseries ALICE} Collaboration, K.~Aamodt {\em et~al.}, ``{Higher harmonic
  anisotropic flow measurements of charged particles in Pb-Pb collisions at
  $\sqrt{s_{NN}}$=2.76 TeV},''
  \href{http://dx.doi.org/10.1103/PhysRevLett.107.032301}{{\em Phys.Rev.Lett.}
  {\bfseries 107} (2011) 032301},
\href{http://arxiv.org/abs/arXiv:1105.3865}{{\ttfamily arXiv:arXiv:1105.3865
  [nucl-ex]}}.

\bibitem{Aamodt:2011by}
{\bfseries ALICE} Collaboration, K.~Aamodt {\em et~al.}, ``{Harmonic
  decomposition of two-particle angular correlations in Pb-Pb collisions at
  $\sqrt{s_{NN}}=2.76$ TeV},''
  \href{http://dx.doi.org/10.1016/j.physletb.2012.01.060}{{\em Phys.Lett.}
  {\bfseries B708} (2012) 249--264},
\href{http://arxiv.org/abs/arXiv:1109.2501}{{\ttfamily arXiv:arXiv:1109.2501
  [nucl-ex]}}.

\bibitem{ATLAS:2012at}
{\bfseries ATLAS} Collaboration, G.~Aad {\em et~al.}, ``{Measurement of the
  azimuthal anisotropy for charged particle production in $\sqrt{s_{NN}}=2.76$
  TeV lead-lead collisions with the ATLAS detector},''
  \href{http://dx.doi.org/10.1103/PhysRevC.86.014907}{{\em Phys.Rev.}
  {\bfseries C86} (2012) 014907},
\href{http://arxiv.org/abs/arXiv:1203.3087}{{\ttfamily arXiv:arXiv:1203.3087
  [hep-ex]}}.

\bibitem{Chatrchyan:2012wg}
{\bfseries CMS} Collaboration, S.~Chatrchyan {\em et~al.}, ``{Centrality
  dependence of dihadron correlations and azimuthal anisotropy harmonics in
  PbPb collisions at $\sqrt{s_{NN}}=2.76$ TeV},''
  \href{http://dx.doi.org/10.1140/epjc/s10052-012-2012-3}{{\em Eur.Phys.J.}
  {\bfseries C72} (2012) 2012},
\href{http://arxiv.org/abs/arXiv:1201.3158}{{\ttfamily arXiv:arXiv:1201.3158
  [nucl-ex]}}.

\bibitem{Adamczyk:2013waa}
{\bfseries STAR} Collaboration, L.~Adamczyk {\em et~al.}, ``{Third Harmonic
  Flow of Charged Particles in Au+Au Collisions at sqrtsNN = 200 GeV},''
  \href{http://dx.doi.org/10.1103/PhysRevC.88.014904}{{\em Phys.Rev.}
  {\bfseries C88} (2013) 014904},
\href{http://arxiv.org/abs/arXiv:1301.2187}{{\ttfamily arXiv:arXiv:1301.2187
  [nucl-ex]}}.

\bibitem{CMS:2012qk}
{\bfseries CMS} Collaboration, S.~Chatrchyan {\em et~al.}, ``{Observation of
  long-range near-side angular correlations in proton-lead collisions at the
  LHC},'' \href{http://dx.doi.org/10.1016/j.physletb.2012.11.025}{{\em
  Phys.Lett.} {\bfseries B718} (2013) 795--814},
\href{http://arxiv.org/abs/arXiv:1210.5482}{{\ttfamily arXiv:arXiv:1210.5482
  [nucl-ex]}}.

\bibitem{Abelev:2012ola}
{\bfseries ALICE} Collaboration, B.~Abelev {\em et~al.}, ``{Long-range angular
  correlations on the near and away side in $p$-Pb collisions at
  $\sqrt{s_{NN}}=5.02$ TeV},''
  \href{http://dx.doi.org/10.1016/j.physletb.2013.01.012}{{\em Phys.Lett.}
  {\bfseries B719} (2013) 29--41},
\href{http://arxiv.org/abs/arXiv:1212.2001}{{\ttfamily arXiv:arXiv:1212.2001}}.

\bibitem{Aad:2012gla}
{\bfseries ATLAS} Collaboration, G.~Aad {\em et~al.}, ``{Observation of
  Associated Near-side and Away-side Long-range Correlations in
  $\sqrt{s_{NN}}$=5.02 TeV Proton-lead Collisions with the ATLAS Detector},''
  \href{http://dx.doi.org/10.1103/PhysRevLett.110.182302}{{\em Phys.Rev.Lett.}
  {\bfseries 110} (2013) 182302},
\href{http://arxiv.org/abs/arXiv:1212.5198}{{\ttfamily arXiv:arXiv:1212.5198
  [hep-ex]}}.

\bibitem{Aad:2013fja}
{\bfseries ATLAS} Collaboration, G.~Aad {\em et~al.}, ``{Measurement with the
  ATLAS detector of multi-particle azimuthal correlations in p+Pb collisions at
  $\sqrt{s_{NN}}$=5.02 TeV},''
  \href{http://dx.doi.org/10.1016/j.physletb.2013.06.057}{{\em Phys.Lett.}
  {\bfseries B725} (2013) 60--78},
\href{http://arxiv.org/abs/1303.2084}{{\ttfamily arXiv:1303.2084 [hep-ex]}}.

\bibitem{Chatrchyan:2013nka}
{\bfseries CMS} Collaboration, S.~Chatrchyan {\em et~al.}, ``{Multiplicity and
  transverse momentum dependence of two- and four-particle correlations in pPb
  and PbPb collisions},''
  \href{http://dx.doi.org/10.1016/j.physletb.2013.06.028}{{\em Phys.Lett.}
  {\bfseries B724} (2013) 213--240},
\href{http://arxiv.org/abs/arXiv:1305.0609}{{\ttfamily arXiv:arXiv:1305.0609
  [nucl-ex]}}.

\bibitem{Abelev:2012di}
{\bfseries ALICE} Collaboration, B.~Abelev {\em et~al.}, ``{Anisotropic flow of
  charged hadrons, pions and (anti-)protons measured at high transverse
  momentum in Pb-Pb collisions at $\sqrt{s_{NN}}$=2.76 TeV},''
  \href{http://dx.doi.org/10.1016/j.physletb.2012.12.066}{{\em Phys.Lett.}
  {\bfseries B719} (2013) 18--28},
\href{http://arxiv.org/abs/arXiv:1205.5761}{{\ttfamily arXiv:arXiv:1205.5761
  [nucl-ex]}}.

\bibitem{Abelev:2013haa}
{\bfseries ALICE} Collaboration, B.~B. Abelev {\em et~al.}, ``{Multiplicity
  Dependence of Pion, Kaon, Proton and Lambda Production in p-Pb Collisions at
  $\sqrt(s_NN)$ = 5.02 TeV},''
  \href{http://dx.doi.org/10.1016/j.physletb.2013.11.020}{{\em Phys.Lett.}
  {\bfseries B728} (2014) 25--38},
\href{http://arxiv.org/abs/1307.6796}{{\ttfamily arXiv:1307.6796 [nucl-ex]}}.

\bibitem{Bozek:2012gr}
P.~Bozek and W.~Broniowski, ``{Correlations from hydrodynamic flow in p-Pb
  collisions},'' \href{http://dx.doi.org/10.1016/j.physletb.2012.12.051}{{\em
  Phys.Lett.} {\bfseries B718} (2013) 1557--1561},
\href{http://arxiv.org/abs/arXiv:1211.0845}{{\ttfamily arXiv:arXiv:1211.0845
  [nucl-th]}}.

\bibitem{Dusling:2012wy}
K.~Dusling and R.~Venugopalan, ``{Explanation of systematics of CMS p+Pb high
  multiplicity di-hadron data at $\sqrt{s}_{\rm NN} = 5.02$ TeV},''
  \href{http://dx.doi.org/10.1103/PhysRevD.87.054014}{{\em Phys.Rev.}
  {\bfseries D87} no.~5, (2013) 054014},
\href{http://arxiv.org/abs/arXiv:1211.3701}{{\ttfamily arXiv:arXiv:1211.3701
  [hep-ph]}}.

\bibitem{Dumitru:2013tja}
A.~Dumitru, T.~Lappi, and L.~McLerran, ``{Are the Angular Correlations in pA
  Collisions due to a Glasmion or Bose Condensation ?},''
  \href{http://dx.doi.org/10.1016/j.nuclphysa.2013.12.001}{{\em Nucl.Phys.}
  {\bfseries A922} (2014) 140--149},
\href{http://arxiv.org/abs/arXiv:1310.7136}{{\ttfamily arXiv:arXiv:1310.7136
  [hep-ph]}}.

\bibitem{Basar:2013hea}
G.~Basar and D.~Teaney, ``{A scaling relation between protonÐnucleus and
  nucleusÐnucleus collisions},''
  \href{http://dx.doi.org/10.1016/j.nuclphysa.2014.08.058}{{\em Nucl.Phys.A (In
  Press)} (September 2014) },
\href{http://arxiv.org/abs/arXiv:1312.6770}{{\ttfamily arXiv:arXiv:1312.6770
  [nucl-th]}}.

\bibitem{Borghini:2001vi}
N.~Borghini, P.~M. Dinh, and J.-Y. Ollitrault, ``{Flow analysis from
  multiparticle azimuthal correlations},''
  \href{http://dx.doi.org/10.1103/PhysRevC.64.054901}{{\em Phys.Rev.}
  {\bfseries C64} (2001) 054901},
\href{http://arxiv.org/abs/nucl-th/0105040}{{\ttfamily arXiv:nucl-th/0105040
  [nucl-th]}}.

\bibitem{Bilandzic:2010jr}
A.~Bilandzic, R.~Snellings, and S.~Voloshin, ``{Flow analysis with cumulants:
  Direct calculations},''
  \href{http://dx.doi.org/10.1103/PhysRevC.83.044913}{{\em Phys.Rev.}
  {\bfseries C83} (2011) 044913},
\href{http://arxiv.org/abs/arXiv:1010.0233}{{\ttfamily arXiv:arXiv:1010.0233
  [nucl-ex]}}.

\bibitem{Voloshin:2008dg}
S.~A. Voloshin, A.~M. Poskanzer, and R.~Snellings, ``{Collective phenomena in
  non-central nuclear collisions},'' {\em Relativistic Heavy Ion Physics,
  Landolt-Bšrnstein - Group I Elementary Particles, Nuclei and Atoms}
  {\bfseries 23} (2008) 293--333,
\href{http://arxiv.org/abs/arXiv:0809.2949}{{\ttfamily arXiv:arXiv:0809.2949
  [nucl-ex]}}.

\bibitem{AnteT}
A.~Bilandzic, {\em Anisotropic Flow Measurements in ALICE at the Large Hadron
  Collider}.
\newblock Ph.d thesis, Utrecht University, 2011.
\newblock
  \url{http://www.nikhef.nl/pub/services/biblio/theses_pdf/thesis_A_Bilandzic.pdf}.

\bibitem{Aamodt:2008zz}
{\bfseries ALICE} Collaboration, K.~Aamodt {\em et~al.}, ``The alice experiment
  at the cern lhc,''
\href{http://dx.doi.org/10.1088/1748-0221/3/08/S08002}{{\em Journal of
  Instrumentation} {\bfseries 3} (2008) S08002}.

\bibitem{Abelev:2012ej}
{\bfseries ALICE} Collaboration, B.~Abelev {\em et~al.}, ``{Measurement of
  Event Background Fluctuations for Charged Particle Jet Reconstruction in
  Pb-Pb collisions at $\sqrt{s_{NN}} = 2.76$ TeV},''
  \href{http://dx.doi.org/10.1007/JHEP03(2012)053}{{\em JHEP} {\bfseries 1203}
  (2012) 053},
\href{http://arxiv.org/abs/1201.2423}{{\ttfamily arXiv:1201.2423 [hep-ex]}}.

\bibitem{Wang:1991hta}
X.-N. Wang and M.~Gyulassy, ``{HIJING: A Monte Carlo model for multiple jet
  production in p p, p A and A A collisions},''
\href{http://dx.doi.org/10.1103/PhysRevD.44.3501}{{\em Phys.Rev.} {\bfseries
  D44} (1991) 3501--3516}.

\bibitem{Roesler:2000he}
S.~Roesler, R.~Engel, and J.~Ranft, ``The monte carlo event generator
  dpmjet-iii,'' {\em Proceedings for the International Conference on Advanced
  Monte Carlo for Radiation Physics, Particle Transport Simulation and
  Applications (MC 2000)} (2000) 1033--1038,
\href{http://arxiv.org/abs/hep-ph/0012252}{{\ttfamily arXiv:hep-ph/0012252
  [hep-ph]}}.

\bibitem{Brun:1994aa}
R.~Brun, F.~Carminati, and S.~Giani, ``{GEANT Detector Description and
  Simulation Tool},'' 1994.
\newblock \url{http://wwwinfo.cern.ch/asdoc/psdir/geant/geantall.ps.gz}.

\bibitem{ALICE:2005aa}
{\bfseries ALICE} Collaboration, ``{ALICE technical design report of the
  computing},'' { \it CERN-LHCC-2005-018}, 2005.
\newblock \url{https://cds.cern.ch/record/832753/files/ALICE-TDR-012.pdf}.

\bibitem{Lin:2004en}
Z.-W. Lin, C.~M. Ko, B.-A. Li, B.~Zhang, and S.~Pal, ``A multi-phase transport
  model for relativistic heavy ion collisions,''
  \href{http://dx.doi.org/10.1103/PhysRevC.72.064901}{{\em Phys.Rev.}
  {\bfseries C72} (2005) 064901},
\href{http://arxiv.org/abs/nucl-th/0411110}{{\ttfamily arXiv:nucl-th/0411110
  [nucl-th]}}.

\bibitem{Ma:2014pva}
G.-L. Ma and A.~Bzdak, ``{Long-range azimuthal correlations in proton-proton
  and proton-nucleus collisions from the incoherent scattering of partons},''
  {\em arXiv:1404.4129} (2014) ,
\href{http://arxiv.org/abs/1404.4129}{{\ttfamily arXiv:1404.4129 [hep-ph]}}.

\bibitem{Yan:2013laa}
L.~Yan and J.-Y. Ollitrault, ``{Universal fluctuation-driven eccentricities in
  proton-proton, proton-nucleus and nucleus-nucleus collisions},''
  \href{http://dx.doi.org/10.1103/PhysRevLett.112.082301}{{\em Phys.Rev.Lett.}
  {\bfseries 112} (2014) 082301},
\href{http://arxiv.org/abs/1312.6555}{{\ttfamily arXiv:1312.6555 [nucl-th]}}.

\bibitem{Voloshin:2007pc}
S.~A. Voloshin, A.~M. Poskanzer, A.~Tang, and G.~Wang, ``{Elliptic flow in the
  Gaussian model of eccentricity fluctuations},''
  \href{http://dx.doi.org/10.1016/j.physletb.2007.11.043}{{\em Phys.Lett.}
  {\bfseries B659} (2008) 537--541},
\href{http://arxiv.org/abs/arXiv:0708.0800}{{\ttfamily arXiv:arXiv:0708.0800
  [nucl-th]}}.

\bibitem{Bzdak:2013rya}
A.~Bzdak, P.~Bozek, and L.~McLerran, ``{Fluctuation induced equality of
  multi-particle eccentricities for four or more particles},''
  \href{http://dx.doi.org/10.1016/j.nuclphysa.2014.03.007}{{\em Nucl.Phys.}
  {\bfseries A927} (2014) 15--23},
\href{http://arxiv.org/abs/1311.7325}{{\ttfamily arXiv:1311.7325 [hep-ph]}}.

\bibitem{Agakishiev:2011eq}
{\bfseries STAR} Collaboration, G.~Agakishiev {\em et~al.}, ``{Energy and
  system-size dependence of two- and four-particle $v_2$ measurements in
  heavy-ion collisions at RHIC and their implications on flow fluctuations and
  nonflow},'' \href{http://dx.doi.org/10.1103/PhysRevC.86.014904}{{\em
  Phys.Rev.} {\bfseries C86} (2012) 014904},
\href{http://arxiv.org/abs/1111.5637}{{\ttfamily arXiv:1111.5637 [nucl-ex]}}.

\bibitem{Wang:2014}
{\bfseries STAR} Collaboration, H.~Wang and P.~Sorensen, ``{Azimuthal
  anisotropy in U+U collisions at STAR},'' {\em arXiv:1406.7522} (2014) ,
\href{http://arxiv.org/abs/1406.7522}{{\ttfamily arXiv:1406.7522 [nucl-ex]}}.

\bibitem{Timmins:2013hq}
{\bfseries ALICE} Collaboration, A.~R. Timmins, ``{Event by event di-hadron
  correlations in Pb-Pb 2.76 TeV collisions from the ALICE experiment},''
  \href{http://dx.doi.org/10.1088/1742-6596/446/1/012031}{{\em
  J.Phys.Conf.Ser.} {\bfseries 446} (2013) 012031},
\href{http://arxiv.org/abs/arXiv:1301.6084}{{\ttfamily arXiv:arXiv:1301.6084
  [nucl-ex]}}.

\bibitem{Aad:2013xma}
{\bfseries ATLAS} Collaboration, G.~Aad {\em et~al.}, ``{Measurement of the
  distributions of event-by-event flow harmonics in lead-lead collisions at =
  2.76 TeV with the ATLAS detector at the LHC},''
  \href{http://dx.doi.org/10.1007/JHEP11(2013)183}{{\em JHEP} {\bfseries 1311}
  (2013) 183},
\href{http://arxiv.org/abs/1305.2942}{{\ttfamily arXiv:1305.2942 [hep-ex]}}.

\bibitem{Alver:2010rt}
{\bfseries PHOBOS} Collaboration, B.~Alver {\em et~al.}, ``{Non-flow
  correlations and elliptic flow fluctuations in gold-gold collisions at
  $\sqrt{s_{NN}}=200$ GeV},''
  \href{http://dx.doi.org/10.1103/PhysRevC.81.034915}{{\em Phys.Rev.}
  {\bfseries C81} (2010) 034915},
\href{http://arxiv.org/abs/1002.0534}{{\ttfamily arXiv:1002.0534 [nucl-ex]}}.

\bibitem{Broniowski:2007ft}
W.~Broniowski, P.~Bozek, and M.~Rybczynski, ``{Fluctuating initial conditions
  in heavy-ion collisions from the Glauber approach},''
  \href{http://dx.doi.org/10.1103/PhysRevC.76.054905}{{\em Phys.Rev.}
  {\bfseries C76} (2007) 054905},
\href{http://arxiv.org/abs/0706.4266}{{\ttfamily arXiv:0706.4266 [nucl-th]}}.

\end{thebibliography}\endgroup

\appendix
\section{Tables}
\begin{table*}[th!f]
\begin{center}
\begin{tabular}{c|c|c|c}
{ \bf Uncorrected} & { \bf Corrected } & { \bf Fractional of hadronic cross section }&  { \bf  Fraction of hadronic cross section   }\\
{ \bf $N_{\rm ch}$ bin } & { \bf $\langle N_{\rm ch} \rangle $} & { \bf within bin } & { \bf above lower bin edge } \\
\hline
$[6,12]$ & 12.0 & 0.154 & 0.826  \\
$[12,18]$ & 19.5 & 0.138 & 0.673  \\
$[18,24]$ & 27.1 & 0.122 & 0.535  \\
$[24,30]$ & 34.6 & 0.105 & 0.412  \\
$[30,40]$ & 44.3 & 0.132 & 0.308  \\
$[40,50]$ & 56.8 & 0.0836 & 0.176  \\
$[50,60]$ & 69.2 & 0.0477 & 0.0921  \\
$[60,70]$ & 81.6 & 0.0245 & 0.0444  \\
$[70,80]$ & 94.1 & 0.0116 & 0.0199  \\
$[80,100]$ & 110 & 0.00712 & 0.00831  \\
$[100,120]$ & 135 & 0.00106 & 0.00120  \\
$[120,140]$ & 159 & 0.00012 & 0.00014  \\
$[140,180]$ & 186 & 0.00001 & 0.00001 
\end{tabular}
\caption{Relation of charged track multiplicity $N_{\rm ch}$ to the fraction of hadronic cross section in \pPb collisions. There is a 3.5\% uncertainty in the cross section values. $N_{ch}$ corresponds to the number of charged tracks with $0.2 < p_T < 3$ GeV/$c$ and $|\eta| < 1$. The corrected values of $N_{\rm ch}$ have a systematic uncertainty of 6.0\%}
\label{tab:frXpB}\label{t2}
\end{center}
\end{table*}

\begin{table*}[tbh!f]
\begin{center}
\begin{tabular}{c|c|c|c}
{ \bf Uncorrected} & { \bf Corrected } & { \bf Fraction of hadronic cross section }&  { \bf Fraction of hadronic cross section } \\
{ \bf $N_{\rm ch}$ bin} & { \bf $\langle N_{\rm ch} \rangle$ } & { \bf within bin } & { \bf above lower bin edge } \\
\hline
$[6,26]$ & 19.82 & 0.111 & 0.928 \\ 
$[26,46]$ & 46.7 & 0.0616 & 0.817 \\
$[46,76]$ & 79.0 & 0.0615 & 0.755 \\
$[76,106]$ & 118 & 0.0446 & 0.694 \\
$[106,150]$ & 166 & 0.0504 & 0.649 \\
$[150,200]$ & 227 & 0.0453 & 0.599 \\
$[200,250]$ & 292 & 0.0377 & 0.553 \\
$[250,300]$ & 358 & 0.0326 & 0.516 \\
$[300,350]$ & 423 & 0.0289 & 0.483 \\
$[350,400]$ & 488 & 0.0261 & 0.454 \\
$[400,450]$ & 552 & 0.0238 & 0.428 \\
$[450,500]$ & 618 & 0.0221 & 0.404 \\
$[500,600]$ & 714 & 0.0397 & 0.382 \\
$[600,700]$ & 843 & 0.0351 & 0.342 \\
$[700,800]$ & 973 & 0.0316 & 0.307 \\
$[800,900]$ & 1103 & 0.0286 & 0.276 \\
$[900,1000]$ & 1233 & 0.0262 & 0.247 \\
$[1000,1200]$ & 1425 & 0.0466 & 0.221 \\
$[1200,1400]$ & 1684 & 0.0402 & 0.174 \\
$[1400,1600]$ & 1944 & 0.0352 & 0.134 \\
$[1600,1800]$ & 2203 & 0.0307 & 0.0990 \\
$[1800,2000]$ & 2462 & 0.0268 & 0.0683 \\
$[2000,2400]$ & 2819 & 0.0388 & 0.0415 \\
\hline
$[1900,1950]$ & 2497 & 0.00656 & 0.0544 \\
$[1950,2000]$ & 2562 & 0.00635 & 0.0478 \\
$[2000,2050]$ & 2627 & 0.00617 & 0.0415 \\ 
$[2050,2100]$ & 2692 & 0.00594 & 0.0353 \\
$[2100,2150]$ & 2757 & 0.00570 & 0.0293 \\
$[2150,2200]$ & 2822 & 0.00544 & 0.0236 \\
$[2200,2250]$ & 2886 & 0.00502 & 0.0182 \\
$[2250,2300]$ & 2951 & 0.00445 & 0.0132 \\
$[2300,2350]$ & 3015 & 0.00353 & 0.00873 \\
$[2350,2400]$ & 3079 & 0.00249 & 0.00520 \\
$[2400,2450]$ & 3143 & 0.00151 & 0.00271 \\
$[2450,2500]$ & 3206 & 0.00074 & 0.00120 \\
$[2500,2550]$ & 3270 & 0.00031 & 0.00045 \\
$[2550,2600]$ & 3334 & 0.00010 & 0.00014

\end{tabular}
\caption{Relation of charged track multiplicity $N_{\rm ch}$ to the fraction of hadronic cross section in \Pb collisions. There is a 1\% uncertainty in the cross section values. $N_{ch}$ corresponds to the number of charged tracks with $0.2 < p_T < 3$ GeV/$c$ and $|\eta| < 1$. The corrected values of $N_{\rm ch}$ have a systematic uncertainty of 6.0\%.}
\label{t3}
\end{center}
\end{table*}
%
\clearpage
\newpage
\section{The ALICE Collaboration}
\label{app:collab}



\begingroup
\small
\begin{flushleft}
B.~Abelev\Irefn{org71}\And
J.~Adam\Irefn{org37}\And
D.~Adamov\'{a}\Irefn{org79}\And
M.M.~Aggarwal\Irefn{org83}\And
G.~Aglieri~Rinella\Irefn{org34}\And
M.~Agnello\Irefn{org107}\textsuperscript{,}\Irefn{org90}\And
A.~Agostinelli\Irefn{org26}\And
N.~Agrawal\Irefn{org44}\And
Z.~Ahammed\Irefn{org126}\And
N.~Ahmad\Irefn{org18}
% ...
\renewcommand\labelenumi{\textsuperscript{\theenumi}~}

\section*{Affiliation notes}
\renewcommand\theenumi{\roman{enumi}}
\begin{Authlist}
\item \Adef{0}Deceased
\item \Adef{idp1127200}{Also at: St. Petersburg State Polytechnical University}
\item \Adef{idp3043744}{Also at: Department of Applied Physics, Aligarh Muslim University, Aligarh, India}
\item \Adef{idp3724416}{Also at: M.V. Lomonosov Moscow State University, D.V. Skobeltsyn Institute of Nuclear Physics, Moscow, Russia}
\item \Adef{idp3968176}{Also at: University of Belgrade, Faculty of Physics and "Vin\v{c}a" Institute of Nuclear Sciences, Belgrade, Serbia}
\item \Adef{idp4279792}{Permanent Address: Permanent Address: Konkuk University, Seoul, Korea}
\item \Adef{idp4831104}{Also at: Institute of Theoretical Physics, University of Wroclaw, Wroclaw, Poland}
\item \Adef{idp5750816}{Also at: University of Kansas, Lawrence, KS, United States}
\end{Authlist}

\section*{Collaboration Institutes}
\renewcommand\theenumi{\arabic{enumi}~}
\begin{Authlist}

\item \Idef{org1}A.I. Alikhanyan National Science Laboratory (Yerevan Physics Institute) Foundation, Yerevan, Armenia
\item \Idef{org2}Benem\'{e}rita Universidad Aut\'{o}noma de Puebla, Puebla, Mexico
\item \Idef{org3}Bogolyubov Institute for Theoretical Physics, Kiev, Ukraine
\item \Idef{org4}Bose Institute, Department of Physics and Centre for Astroparticle Physics and Space Science (CAPSS), Kolkata, India
\item \Idef{org5}Budker Institute for Nuclear Physics, Novosibirsk, Russia
\item \Idef{org6}California Polytechnic State University, San Luis Obispo, CA, United States
\item \Idef{org7}Central China Normal University, Wuhan, China
\item \Idef{org8}Centre de Calcul de l'IN2P3, Villeurbanne, France
\item \Idef{org9}Centro de Aplicaciones Tecnol\'{o}gicas y Desarrollo Nuclear (CEADEN), Havana, Cuba
\item \Idef{org10}Centro de Investigaciones Energ\'{e}ticas Medioambientales y Tecnol\'{o}gicas (CIEMAT), Madrid, Spain
\item \Idef{org11}Centro de Investigaci\'{o}n y de Estudios Avanzados (CINVESTAV), Mexico City and M\'{e}rida, Mexico
\item \Idef{org12}Centro Fermi - Museo Storico della Fisica e Centro Studi e Ricerche ``Enrico Fermi'', Rome, Italy
\item \Idef{org13}Chicago State University, Chicago, USA
\item \Idef{org14}Commissariat \`{a} l'Energie Atomique, IRFU, Saclay, France
\item \Idef{org15}COMSATS Institute of Information Technology (CIIT), Islamabad, Pakistan
\item \Idef{org16}Departamento de F\'{\i}sica de Part\'{\i}culas and IGFAE, Universidad de Santiago de Compostela, Santiago de Compostela, Spain
\item \Idef{org17}Department of Physics and Technology, University of Bergen, Bergen, Norway
\item \Idef{org18}Department of Physics, Aligarh Muslim University, Aligarh, India
\item \Idef{org19}Department of Physics, Ohio State University, Columbus, OH, United States
\item \Idef{org20}Department of Physics, Sejong University, Seoul, South Korea
\item \Idef{org21}Department of Physics, University of Oslo, Oslo, Norway
\item \Idef{org22}Dipartimento di Fisica dell'Universit\`{a} 'La Sapienza' and Sezione INFN Rome, Italy
\item \Idef{org23}Dipartimento di Fisica dell'Universit\`{a} and Sezione INFN, Cagliari, Italy
\item \Idef{org24}Dipartimento di Fisica dell'Universit\`{a} and Sezione INFN, Trieste, Italy
\item \Idef{org25}Dipartimento di Fisica dell'Universit\`{a} and Sezione INFN, Turin, Italy
\item \Idef{org26}Dipartimento di Fisica e Astronomia dell'Universit\`{a} and Sezione INFN, Bologna, Italy
\item \Idef{org27}Dipartimento di Fisica e Astronomia dell'Universit\`{a} and Sezione INFN, Catania, Italy
\item \Idef{org28}Dipartimento di Fisica e Astronomia dell'Universit\`{a} and Sezione INFN, Padova, Italy
\item \Idef{org29}Dipartimento di Fisica `E.R.~Caianiello' dell'Universit\`{a} and Gruppo Collegato INFN, Salerno, Italy
\item \Idef{org30}Dipartimento di Scienze e Innovazione Tecnologica dell'Universit\`{a} del  Piemonte Orientale and Gruppo Collegato INFN, Alessandria, Italy
\item \Idef{org31}Dipartimento Interateneo di Fisica `M.~Merlin' and Sezione INFN, Bari, Italy
\item \Idef{org32}Division of Experimental High Energy Physics, University of Lund, Lund, Sweden
\item \Idef{org33}Eberhard Karls Universit\"{a}t T\"{u}bingen, T\"{u}bingen, Germany
\item \Idef{org34}European Organization for Nuclear Research (CERN), Geneva, Switzerland
\item \Idef{org35}Faculty of Engineering, Bergen University College, Bergen, Norway
\item \Idef{org36}Faculty of Mathematics, Physics and Informatics, Comenius University, Bratislava, Slovakia
\item \Idef{org37}Faculty of Nuclear Sciences and Physical Engineering, Czech Technical University in Prague, Prague, Czech Republic
\item \Idef{org38}Faculty of Science, P.J.~\v{S}af\'{a}rik University, Ko\v{s}ice, Slovakia
\item \Idef{org39}Frankfurt Institute for Advanced Studies, Johann Wolfgang Goethe-Universit\"{a}t Frankfurt, Frankfurt, Germany
\item \Idef{org40}Gangneung-Wonju National University, Gangneung, South Korea
\item \Idef{org41}Gauhati University, Department of Physics, Guwahati, India
\item \Idef{org42}Helsinki Institute of Physics (HIP), Helsinki, Finland
\item \Idef{org43}Hiroshima University, Hiroshima, Japan
\item \Idef{org44}Indian Institute of Technology Bombay (IIT), Mumbai, India
\item \Idef{org45}Indian Institute of Technology Indore, Indore (IITI), India
\item \Idef{org46}Inha University, Incheon, South Korea
\item \Idef{org47}Institut de Physique Nucl\'eaire d'Orsay (IPNO), Universit\'e Paris-Sud, CNRS-IN2P3, Orsay, France
\item \Idef{org48}Institut f\"{u}r Informatik, Johann Wolfgang Goethe-Universit\"{a}t Frankfurt, Frankfurt, Germany
\item \Idef{org49}Institut f\"{u}r Kernphysik, Johann Wolfgang Goethe-Universit\"{a}t Frankfurt, Frankfurt, Germany
\item \Idef{org50}Institut f\"{u}r Kernphysik, Westf\"{a}lische Wilhelms-Universit\"{a}t M\"{u}nster, M\"{u}nster, Germany
\item \Idef{org51}Institut Pluridisciplinaire Hubert Curien (IPHC), Universit\'{e} de Strasbourg, CNRS-IN2P3, Strasbourg, France
\item \Idef{org52}Institute for Nuclear Research, Academy of Sciences, Moscow, Russia
\item \Idef{org53}Institute for Subatomic Physics of Utrecht University, Utrecht, Netherlands
\item \Idef{org54}Institute for Theoretical and Experimental Physics, Moscow, Russia
\item \Idef{org55}Institute of Experimental Physics, Slovak Academy of Sciences, Ko\v{s}ice, Slovakia
\item \Idef{org56}Institute of Physics, Academy of Sciences of the Czech Republic, Prague, Czech Republic
\item \Idef{org57}Institute of Physics, Bhubaneswar, India
\item \Idef{org58}Institute of Space Science (ISS), Bucharest, Romania
\item \Idef{org59}Instituto de Ciencias Nucleares, Universidad Nacional Aut\'{o}noma de M\'{e}xico, Mexico City, Mexico
\item \Idef{org60}Instituto de F\'{\i}sica, Universidad Nacional Aut\'{o}noma de M\'{e}xico, Mexico City, Mexico
\item \Idef{org61}iThemba LABS, National Research Foundation, Somerset West, South Africa
\item \Idef{org62}Joint Institute for Nuclear Research (JINR), Dubna, Russia
\item \Idef{org63}Konkuk University, Seoul, South Korea
\item \Idef{org64}Korea Institute of Science and Technology Information, Daejeon, South Korea
\item \Idef{org65}KTO Karatay University, Konya, Turkey
\item \Idef{org66}Laboratoire de Physique Corpusculaire (LPC), Clermont Universit\'{e}, Universit\'{e} Blaise Pascal, CNRS--IN2P3, Clermont-Ferrand, France
\item \Idef{org67}Laboratoire de Physique Subatomique et de Cosmologie, Universit\'{e} Grenoble-Alpes, CNRS-IN2P3, Grenoble, France
\item \Idef{org68}Laboratori Nazionali di Frascati, INFN, Frascati, Italy
\item \Idef{org69}Laboratori Nazionali di Legnaro, INFN, Legnaro, Italy
\item \Idef{org70}Lawrence Berkeley National Laboratory, Berkeley, CA, United States
\item \Idef{org71}Lawrence Livermore National Laboratory, Livermore, CA, United States
\item \Idef{org72}Moscow Engineering Physics Institute, Moscow, Russia
\item \Idef{org73}National Centre for Nuclear Studies, Warsaw, Poland
\item \Idef{org74}National Institute for Physics and Nuclear Engineering, Bucharest, Romania
\item \Idef{org75}National Institute of Science Education and Research, Bhubaneswar, India
\item \Idef{org76}Niels Bohr Institute, University of Copenhagen, Copenhagen, Denmark
\item \Idef{org77}Nikhef, National Institute for Subatomic Physics, Amsterdam, Netherlands
\item \Idef{org78}Nuclear Physics Group, STFC Daresbury Laboratory, Daresbury, United Kingdom
\item \Idef{org79}Nuclear Physics Institute, Academy of Sciences of the Czech Republic, \v{R}e\v{z} u Prahy, Czech Republic
\item \Idef{org80}Oak Ridge National Laboratory, Oak Ridge, TN, United States
\item \Idef{org81}Petersburg Nuclear Physics Institute, Gatchina, Russia
\item \Idef{org82}Physics Department, Creighton University, Omaha, NE, United States
\item \Idef{org83}Physics Department, Panjab University, Chandigarh, India
\item \Idef{org84}Physics Department, University of Athens, Athens, Greece
\item \Idef{org85}Physics Department, University of Cape Town, Cape Town, South Africa
\item \Idef{org86}Physics Department, University of Jammu, Jammu, India
\item \Idef{org87}Physics Department, University of Rajasthan, Jaipur, India
\item \Idef{org88}Physik Department, Technische Universit\"{a}t M\"{u}nchen, Munich, Germany
\item \Idef{org89}Physikalisches Institut, Ruprecht-Karls-Universit\"{a}t Heidelberg, Heidelberg, Germany
\item \Idef{org90}Politecnico di Torino, Turin, Italy
\item \Idef{org91}Purdue University, West Lafayette, IN, United States
\item \Idef{org92}Pusan National University, Pusan, South Korea
\item \Idef{org93}Research Division and ExtreMe Matter Institute EMMI, GSI Helmholtzzentrum f\"ur Schwerionenforschung, Darmstadt, Germany
\item \Idef{org94}Rudjer Bo\v{s}kovi\'{c} Institute, Zagreb, Croatia
\item \Idef{org95}Russian Federal Nuclear Center (VNIIEF), Sarov, Russia
\item \Idef{org96}Russian Research Centre Kurchatov Institute, Moscow, Russia
\item \Idef{org97}Saha Institute of Nuclear Physics, Kolkata, India
\item \Idef{org98}School of Physics and Astronomy, University of Birmingham, Birmingham, United Kingdom
\item \Idef{org99}Secci\'{o}n F\'{\i}sica, Departamento de Ciencias, Pontificia Universidad Cat\'{o}lica del Per\'{u}, Lima, Peru
\item \Idef{org100}Sezione INFN, Bari, Italy
\item \Idef{org101}Sezione INFN, Bologna, Italy
\item \Idef{org102}Sezione INFN, Cagliari, Italy
\item \Idef{org103}Sezione INFN, Catania, Italy
\item \Idef{org104}Sezione INFN, Padova, Italy
\item \Idef{org105}Sezione INFN, Rome, Italy
\item \Idef{org106}Sezione INFN, Trieste, Italy
\item \Idef{org107}Sezione INFN, Turin, Italy
\item \Idef{org108}SSC IHEP of NRC Kurchatov institute, Protvino, Russia
\item \Idef{org109}SUBATECH, Ecole des Mines de Nantes, Universit\'{e} de Nantes, CNRS-IN2P3, Nantes, France
\item \Idef{org110}Suranaree University of Technology, Nakhon Ratchasima, Thailand
\item \Idef{org111}Technical University of Split FESB, Split, Croatia
\item \Idef{org112}The Henryk Niewodniczanski Institute of Nuclear Physics, Polish Academy of Sciences, Cracow, Poland
\item \Idef{org113}The University of Texas at Austin, Physics Department, Austin, TX, USA
\item \Idef{org114}Universidad Aut\'{o}noma de Sinaloa, Culiac\'{a}n, Mexico
\item \Idef{org115}Universidade de S\~{a}o Paulo (USP), S\~{a}o Paulo, Brazil
\item \Idef{org116}Universidade Estadual de Campinas (UNICAMP), Campinas, Brazil
\item \Idef{org117}University of Houston, Houston, TX, United States
\item \Idef{org118}University of Jyv\"{a}skyl\"{a}, Jyv\"{a}skyl\"{a}, Finland
\item \Idef{org119}University of Liverpool, Liverpool, United Kingdom
\item \Idef{org120}University of Tennessee, Knoxville, TN, United States
\item \Idef{org121}University of Tokyo, Tokyo, Japan
\item \Idef{org122}University of Tsukuba, Tsukuba, Japan
\item \Idef{org123}University of Zagreb, Zagreb, Croatia
\item \Idef{org124}Universit\'{e} de Lyon, Universit\'{e} Lyon 1, CNRS/IN2P3, IPN-Lyon, Villeurbanne, France
\item \Idef{org125}V.~Fock Institute for Physics, St. Petersburg State University, St. Petersburg, Russia
\item \Idef{org126}Variable Energy Cyclotron Centre, Kolkata, India
\item \Idef{org127}Vestfold University College, Tonsberg, Norway
\item \Idef{org128}Warsaw University of Technology, Warsaw, Poland
\item \Idef{org129}Wayne State University, Detroit, MI, United States
\item \Idef{org130}Wigner Research Centre for Physics, Hungarian Academy of Sciences, Budapest, Hungary
\item \Idef{org131}Yale University, New Haven, CT, United States
\item \Idef{org132}Yonsei University, Seoul, South Korea
\item \Idef{org133}Zentrum f\"{u}r Technologietransfer und Telekommunikation (ZTT), Fachhochschule Worms, Worms, Germany
\end{Authlist}
\endgroup

\end{document}